%% file: dis99.tex
\documentstyle[twoside,fleqn,npb,epsfig,rotating]{article}
\input abbrevia

\begin{document}
%
%
\title{\vspace{-1.2cm} 
{\normalsize\rightline{MPI-PhE/99-04}
}
Latest Highlights from the H1 Collaboration}
\author{Tancredi Carli\address{Max-Planck Institut f\"ur Physik, \\
        F\"ohringer Ring 6, 80805 M\"unchen}%
        \thanks{Invited plenary talk at VII. Int. workshop on
DIS and QCD, April 1999, Zeuthen-Berlin
        }
}
\begin{abstract}
Recent data obtained in $e^+ p$ collisions at HERA by the
H1 collaboration are presented: searches for new phenomena
beyond the Standard Model, measurement of the inclusive
neutral and charged current cross sections and of the proton
structure function, determination of the gluon density in the
proton and photon from jet productions, measurement and
interpretation of event shapes in the
current region of the Breit frame, test of the parton evolution
at low $x$ and of the diffractive production mechanism.
\end{abstract}

\maketitle
\input{intro.tex}

%
\input{excess.tex}

\input{discross.tex}
\input{f2sum.tex}

\input{jets.tex}
\input{gluonpho.tex}

%
\input{eventsha.tex}

\input{pi0.tex}

\input{diffchar.tex}

\input{vmmodel.tex}

\section{Conclusions} 
\input{conclusions.tex}

\vspace{-0.35cm}
\section*{Acknowledgments}
\vspace{-0.2cm}
I would like to thank the organisers for this stimulating workshop. 
It is a pleasure to thank my colleagues from the H1 collaboration
for their enthusiasm to produce the results 
which made this report possible.
Many to thanks to M. Erdmann for the critical reading of the
manuscript.
%

\input dis99.bbl
\end{document}

%% file: abbrevia.tex
\newcommand{\av}[1]{\mbox{$ \langle #1 \rangle $}}

\newcommand{\xB}{\mbox{$x$}}  

\newcommand{\Qsq}{\mbox{$Q^2$~}}
\newcommand{\Qsqx}{\mbox{$Q^2$}}

\newcommand{\pt}{\mbox{$P^{\star}_{\! T}$}}

\newcommand{\as}{\mbox{$\alpha_s$}}

\newcommand{\cm}{\mbox{\rm ~cm}~}

\newcommand{\GeV}{\mbox{\rm ~GeV}~}
\newcommand{\GeVx}{\mbox{\rm ~GeV}}

\newcommand{\GeVsq}{\mbox{${\rm ~GeV}^2$}~}
\newcommand{\GeVsqx}{\mbox{${\rm ~GeV}^2$}}

\newcommand{\pbinv}{\mbox{${\rm ~pb^{-1}}$}}
\newcommand{\intlum}{\mbox{$\int {\cal L} \; dt = \; $}}    
\newcommand{\jpsi}{\mbox{$J\!/\psi$}}
\newcommand{\sjpsi}{\mbox{\scriptsize $J\!/\psi$}}

\newcommand{\ee}{\mbox{$e^+e^-$}}

\newcommand{\Wgp}{\mbox{$W_{\! \gamma p}$}}

\newcommand{\mean}[1]{\left< #1 \right>}

\newcommand{\mi}{\mu_{\scriptscriptstyle I}}

\newcommand{\fmean}{\mean{F}}
\newcommand{\fpert}{\fmean^{\rm pert}}
\newcommand{\fpow}{\fmean^{\rm pow}}

\newcommand{\lsim}{\raisebox{-1.5mm}{$\:\stackrel{\textstyle{<}}{\textstyle{\sim}}\:$}}
\newcommand{\gsim}{\raisebox{-0.5mm}{$\stackrel{>}{\scriptstyle{\sim}}$}}

\newcommand{\pom}{\rm I\!P}

\newcommand{\alphapom}{\alpha_{_{\rm I\!P}}}

\newcommand{\xpom}{x_{_{I\!\!P}}}
\newcommand{\mx}{M_{_X}}

\newcommand{\zpom}{z_{_{\rm \pom}}}

\def\invpb{\;{\rm pb^{-1}}}                

%% file: intro.tex
\section{Introduction}
The electron-proton collider HERA offers the
unique possibility to explore the complex structure
of the proton in a relatively clean environment.
The large centre of mass energy of $\sqrt{s} \approx 300$~\GeV
allows a perturbative QCD analysis in a previously
inaccessible domain. Compared to fixed target experiments
HERA extends the kinematic reach of deep-inelastic scattering (DIS) 
by about of two orders of magnitude in the Bjorken variable $x$ 
and the photon virtuality $\Qsqx$.
%
The understanding of the proton structure is in many aspects 
a challenge to theory and therefore interesting in itself. 
Furthermore, the proton is the particle which gives access to 
the highest possible centre of mass energy in particle collisions.
The knowledge of the (non-perturbative) parton density
function in the proton and the theoretical understanding
how they evolve in the different kinematic regimes
is therefore an important prerequisite for any possible future
discovery of new phenomena, e.g. 
in proton proton collisions at LHC.

Within the Standard Model many aspects of the strong interactions
remain unexplained. HERA has here the peerless opportunity
to link QCD to the general analytic properties of the 
scattering amplitudes as e.g. exploited in the Regge phenomenology.
Such theories have been very sucessful to describe 
elastic and diffractive
scattering at high energies and the energy dependences of
the total hadronic cross section. First steps to understand
the dynamics of the transformation of a complex ´coloured´ 
system in a colour singlet probed by a virtual photon
have been very sucessful.

In the period $1994-1997$ HERA colliding $E_e = 27.5$ \GeV
positrons on $E_p = 820$ \GeV protons has delivered
an integrated luminosity of about $\intlum 37 \pbinv$
of useful data to H1. Most of the results in this
report will be based on this data set.
Since $1998$ HERA has switched back to electrons and
about $\intlum 15 \pbinv$ have been collected.
First results based on $\intlum 5.6 \pbinv$ from 
electron-proton collisions are already available. 
This represents an increase in $e^- p$ data by a factor of $10$. 
In addition the proton energy was pushed to
$E_p = 920$ \GeV leading to a higher
centre of mass energy and to a higher
parton luminosity at fixed $x$.


%% file: excess.tex
%
\section{Search for New Phenomena at Highest Energy}
%
The high momentum transfers accessible at HERA
allow the proton to be probed
at very small distances down to $10^{-16}$\cm
via exchange of highly virtual gauge
bosons. 
Particles with masses higher than the centre
of mass energy could be exchanged in the $t$-channel.
Their interference with standard DIS processes
could be experimentally observed as enhancement
or deficit in the NC cross section at high $Q^2$
or at high mass.
Moreover, new particles with masses up to $300$ \GeV
could be produced as $s$-channel resonances. 
HERA is here particularly sensitive to particles 
coupling to electrons and quarks in the proton.

\subsection{Inclusive Searches}
\begin{figure}[tbhp]
 \vspace{-1.5cm}
 \mbox{\hspace{-0.5cm}{
 \epsfig{file=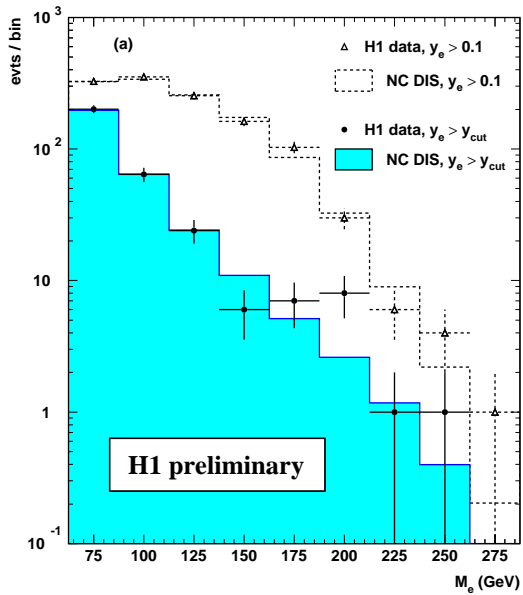,width=8.cm}
 }}
 \vspace{-1.5cm}
 \caption{\it $M = \sqrt{s x}$ distribution for the inclusive
 $e^+ p \to e^+ X$ sample and for a optimised cut in $y$.
 Shown are H1 data and the DIS expectation.}
\label{fig:excess}
 \vspace{-0.8cm}
\end{figure}
In an analysis of the data taken in 
$1994$ to $1996$ an excess of neutral current data 
$e^+ p \to e^+ X $ at very high $Q^2$ has been found.
$12$ events with $\Qsq > 15000$ \GeVsq
have been observed in the data while only
$4.71 \pm 0.76$ were expected by standard DIS.
$7$ of these events had an invariant mass
$M=\sqrt{s x}$ in a window of $25$ \GeV around
$M=200$ \GeV and $y > 0.4$.
In 1997 the total luminosity was doubled. Also in this
data a tendency for more events at high $Q^2$
has been found. The magnitude of the excess is however
not confirmed. Above $\Qsq > 15000$ \GeVsq
$22$ events are found and $14.7 \pm 2.1$
are expected. $8$ events fall in the mass window
at $M=200$ \GeV and $3.01\pm 0.54$ are expected.

The mass distribution of all $e^+ p$
data is shown in Fig. \ref{fig:excess}.
The standard DIS background is known 
within $\pm 5-10 \%$.
If the new particle has spin $0$, the angular
distribution of the decay product is flat.
The $y= 1/2 \, (1+\cos{\theta^*})$ distribution\footnote{
$\theta^*$ is the lepton-parton angle in the centre of mass
system of the decaying particle.}  
will therefore be flat, as opposed to DIS where the
cross section falls like $1/y^2$. 
The signal to background ratio can therefore be 
optimised by a cut in $y$. The mass distribution for
a $y_{cut}(M)$ optimised for each mass bin, is also shown
in Fig.~\ref{fig:excess}.
From this mass spectrum upper limits on scalar and vector 
leptoquarks classified according to their
spin, weak isospin and fermion number~\cite{buchmueller} 
have been derived. If the strength of this 
new $e^+q$ coupling is the same as for the electromagnetic
coupling, than masses up to 275 \GeV are excluded.
If one allows for a branching $LQ \to e q$ of 
e.g. $10\%$, masses up to 255 \GeV are ruled out by the data.
For these small branchings this limit extents far beyond
the domain presently covered in $p\bar{p}$ collisions.
This has triggered intensive searches for other
signatures.
No signal has been found in the reactions
$e^\pm p \to \mu^\pm p$ and $e^\pm p \to \tau^\pm p$.
However, the derived limits on leptoquarks decaying in
the second or third generation
are better as or competitive with low energy experiments
in most cases~\cite{H1vancouverlq}.

\subsection{Muon Events}
In an inclusive search for events 
with a high transverse momentum of the hadronic
final state above $25$ \GeVx,
six events with high energetic isolated leptons 
($\pt > 10 $ \GeVx) have been observed.
One event has an electron and five have a muon in the
final state.
In all events an imbalance of
missing longitudinal and transverse momentum
is measured suggesting the presence of an
undetectable particle like a neutrino. 
The electron event and two muon events are found 
in the phase space where the most important background
- the direct production of $W$ bosons $e p \to W X$ - 
are expected. 
In the muon (electron) channel 
$0.8\pm 0.2$ ($2.4\pm 0.5$) events are expected
and $5$ ($1$) have been observed.
\begin{figure}[tbhp]
 \vspace{-1.5cm}
 \mbox{\hspace{-0.5cm}{
 \epsfig{file=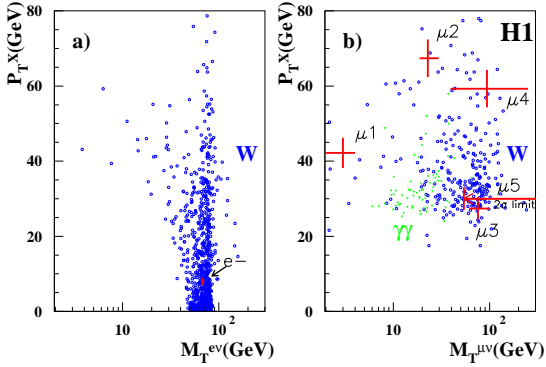,width=8.cm}
 }}
 \vspace{-1.5cm}
 \caption{\it Transverse momentum of the hadronic final state
as a function of the transverse mass of the
lepton-neutrino system for events with isolated leptons.
The crosses gives the $1\sigma$ uncertainty on the measured
parameters. 
The expected distribution for
direct $W^\pm$ production and muon production
in $\gamma \gamma$ processes is overlayed.} 
\label{fig:muon}
 \vspace{-0.8cm}
\end{figure}

The striking characteristics of the muon events
is the large amount of transverse energy of the
hadronic system ($P_T^X$) which is displayed together with
lepton-neutrino transverse mass ($M_T^{\mu \nu}$)
in Fig.~\ref{fig:muon}. Overlayed is 
the expected distribution of direct production of $W$ bosons. 
It would be however interesting to know how QCD
effects which are presently not included in the background
estimation using the event generator EPVEC~\cite{mc:epvec},
could alter the $P_T^X$ and $M_T^{\mu \nu}$ distribution.
In the 1998/1999 $e^- p$ data no events with isolated leptons
have been found. For a luminosity of
$\intlum 5.1 \pbinv$ $0.14 \pm 0.04$ 
($0.37 \pm 0.07$) events are expected
in the muon (electron) channel.

%% file: discross.tex
\section{Inclusive Neutral and Charged Current Single
Differential Cross Sections}
\input{nc99.tex}

\input{cc99.tex}

\subsection{Neutral Current Events}
The Standard Model neutral current cross section
can be represented in the form:
\begin{eqnarray}  
\frac{d^2\sigma^{e^\pm p}}{dx dQ^2} =  
\frac{2 \pi \alpha^2}{x Q^4} 
\left [
Y_+ \tilde{F}_2(x,Q^2) - y^2 F_L(x,Q^2)
\right. \nonumber \\ \left.  
\mp Y_- x F_3(x,Q^2)
\right ] \nonumber 
\end{eqnarray}  
where $\alpha$ is the electromagnetic coupling
constant and
$Y_\pm = 1 \pm (1-y)^2$
contains the helicity dependence of the electroweak interaction.

The generalised proton structure function
$\tilde{F}_2(x,Q^2)$ 
can be decomposed in:
\begin{eqnarray}  
\tilde{F}_2= 
F^{em}_2+ 
\frac{Q^2}{Q^2+M_Z^2}  F^{\gamma Z}_2 +
{\left ( \frac{Q^2}{Q^2+M_Z^2} \right )}^2  F^{Z}_2
\nonumber 
\end{eqnarray}  
where
$F^{em}_2$ describes the pure photon exchange,
$F^{\gamma Z}_2$ the ${\gamma Z}$ interference 
and $F^{Z}_2$ the pure exchange of the $Z$ boson.
$\tilde{F}_2$ is sensitive to the singlet sum
of the quark distributions ($x q + x \bar{q}$). 
$x F_3$ is the parity violating proton structure
function which is sensitive to the non-singlet
difference of the quark densities ($x q - x \bar{q}$)
and is given by:
$$
F_3= 
\frac{Q^2}{Q^2+M_Z^2}  F^{\gamma Z}_3 +
{\left (\frac{Q^2}{Q^2+M_Z^2} \right )}^2  F^{Z}_3
$$
$F_3$ contributes with different sign to the cross
section for the $e^-$ and $e^+$ case.
Below $\Qsq \lsim 1500 \GeVsq$ contributions from 
$Z$ exchange are smaller than $1\% $, but at high \Qsq
the inclusive cross section is sensitive to electroweak effects 
(at $\Qsq = 5000 \GeVsq$ and $x=0.08$: 
$\delta_z - \delta_3  \approx 10 \%$).

The longitudinal structure function $F_L$ 
contributes less than $5 \%$ at very high 
$y$ and
is negligible below $y \le 0.4$. 

The single differential cross section is shown
as a function of \Qsq in Fig. \ref{fig:nc99}
for both $e^+ p$ and $e^- p$ scattering.
The $e^- p$ data have been taken in $1998/1999$
and correspond to the $\int {\cal L} dt \approx 5 \pbinv$ 
analysed up to March $1999$.

The cross section falls like $1/Q^4$ over 
$6$ orders of magnitude and spans $2$ orders of
magnitude in \Qsqx. 
For $\Qsq > 3000$~\GeVsq the $e^-p$ data are always 
above the $e^+ p$ data. 
This provides for the first time evidence for 
$\gamma Z$ interference effects in $e^\pm p$
scattering.
The increase of the centre of
mass energy from $\sqrt{s}=300 $ to $320 $ \GeV 
does not influence this conclusion, since it
leads only to a marginal increase of the cross 
section (see dashed line in Fig.~\ref{fig:nc99}). 
While in the $e^- p$ case good agreement is found
with the Standard Model expectation, in the
$e^+ p$ case a slight overshoot is observed
at the highest \Qsqx.
The theoretical uncertainty mainly introduced
by the parton density functions is about $7\%$
at the highest accessible \Qsq values.

\subsection{Charged Current Events}
Charged current events are characterised by their
missing transverse momentum due to the undetected
neutrino in the final state.
Their cross section is given by:
$$
\frac{d^2\sigma^{e^\pm p}_{CC}}{dx dQ^2} = 
\frac{G^2_F}{2 \pi x} 
{\left (\frac{M^2_{prop}}{Q^2+M_{prop}^2} \right )}^2 
 \cdot \; x \cdot \Phi^{e^\pm p}_{CC}(x,Q^2)
$$
where $G_F$ is the Fermi constant and $M_{prop}$ 
is the mass of the space like charged exchanged boson
which is in the SM the $W$ boson. 
In leading order the helicity weighted parton density
function have the simple form:
$$
 \Phi^{e^+p}_{CC}(x,Q^2)= 
 (\bar{u}+ \bar{c}) + ( 1 - y)^2 \cdot (d+s+b)
$$
$$
 \Phi^{e^-p}_{CC}(x,Q^2)= 
 (u+c) + {( 1 - y)}^2 \cdot (\bar{d}+\bar{s}+\bar{b})
$$
The positrons (electrons) only interact with quarks
or anti-quarks with negative (positive) 
electric charge. 
In the $e^- p$ ($e^+ p$) 
case mostly (anti-)quarks are
involved in the interaction.

Fig.~\ref{fig:cc99} shows the charged current cross
section as a function of $Q^2$. 
The $e^-p$ data are an order of magnitude above $e^+ p$ data.
This is due to the fact that in the $e^- p$ case
the cross section is proportional to $(u+c)$
while for $e^+ p$ the coupling to valence quark
is suppressed by ${(1-y)}^2 \; (d+s)$. 
Moreover is the $d$ quark density smaller than the $u$
quark density in the proton.

The charged current cross section falls over four
orders of magnitudes in \Qsqx. From the \Qsq dependence
the propagator mass has been determined to be:
$$
M_{prop} = 81.2 \pm 3.3 \, (stat.) \pm 4.3 \, (syst) \GeV
$$
This result is compatible with the world average
of the time-like $W$ mass
$M_{W} = 80.41 \pm 0.1 \GeV$ \cite{PDG98}.

At high \Qsq the charged current cross section
is suppressed relative to the neutral current
cross section due to the $d/u$ ratio being less than
unity. From this observation 
it has been determined that at $x=0.3$ 
the $u$ valence quark
density is approximately $1.5 - 3$ times larger than 
$d$ valence quark density.
A determination of $u/d$ ratio at large $x$
free from uncertainties due to nuclear binding
effects will be
possible in the future.

These measurements represent a benchmark for the 
Standard Model of electroweak interactions.
More $e^- p$ and $e^+ p$ data 
at the highest possible \Qsq are needed to get the 
best sensitivity.
HERA will increase the luminosity in the year $2000$.
High statistics data will then allow a better 
understanding of the detector and decisive tests
for deviation from the standard model are possible.
This physics is just starting at HERA
and  a surprise is possible!

%% file: nc99.tex
%
\begin{figure}[htp] 
\vspace{-0.6cm}
 \begin{center}
\mbox{
\hspace{-0.0cm}{
\epsfig{file=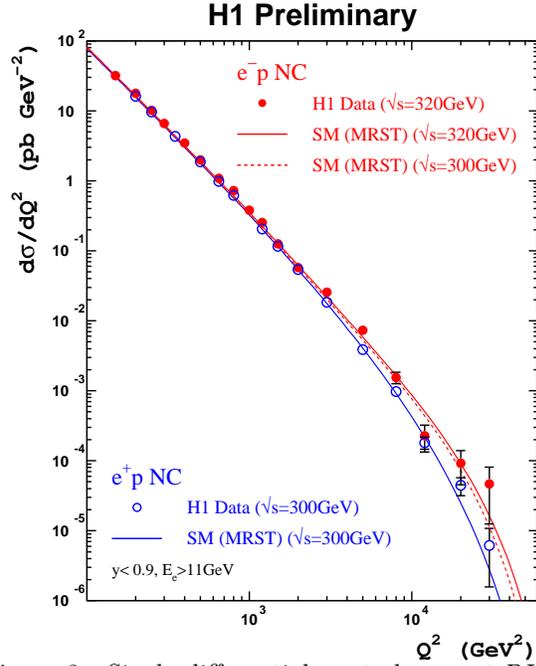,
bbllx=11,bblly=116,bburx=548,bbury=697,clip,width=8.0cm}
}}
 \end{center}
\vspace{-1.5cm}
 \caption{\it Single differential neutral current DIS cross section
 as a function of $Q^2$. 
}
\label{fig:nc99}
\end{figure}
%

%% file: cc99.tex
%
\begin{figure}[htp] 
 \begin{center}
\vspace{-1.4cm}
\mbox{
\hspace{-0.4cm}{
   \epsfig{file=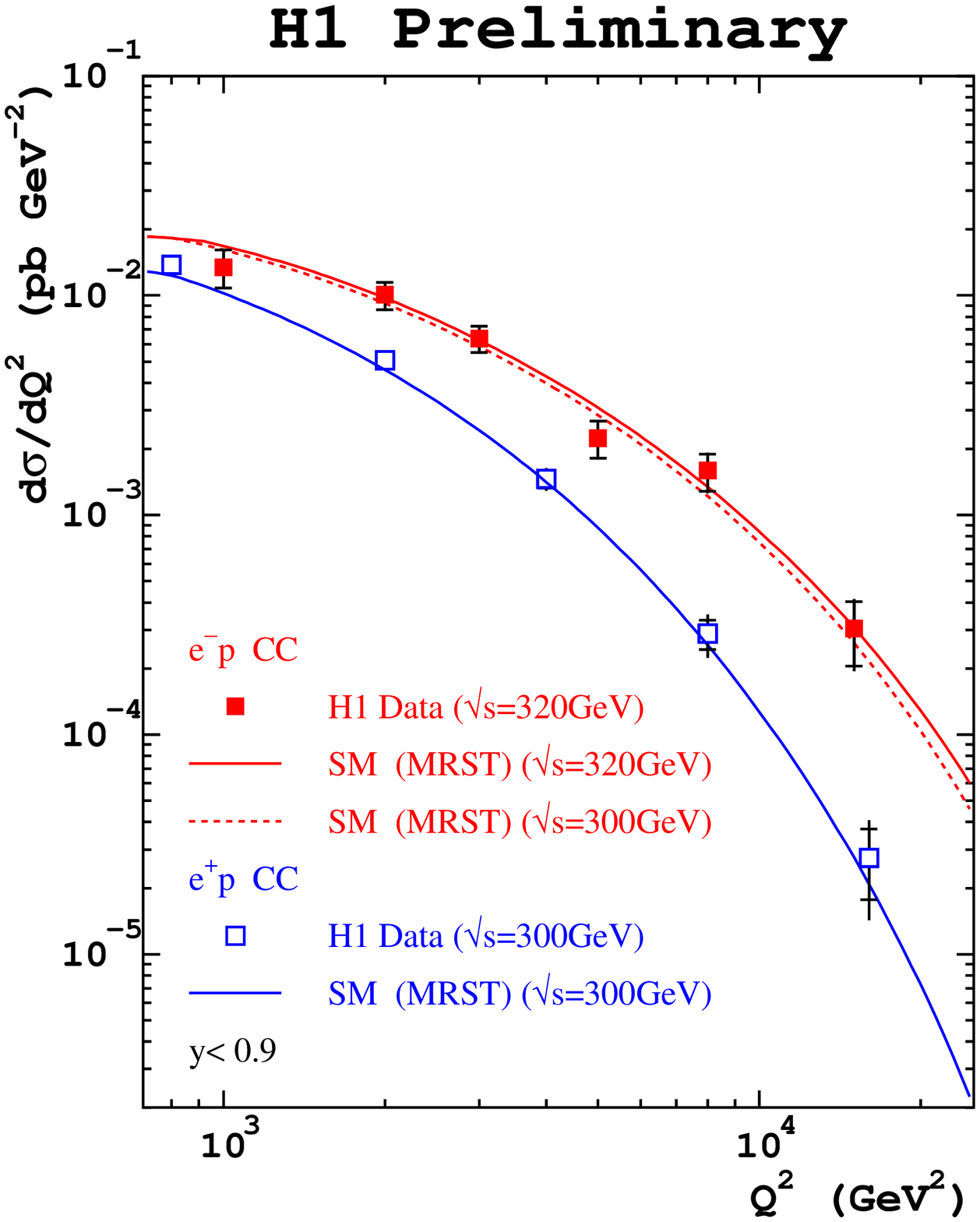,
   bbllx=14,bblly=107,bburx=514,bbury=718,clip,width=8.cm}
}}
\vspace{-1.4cm}
 \caption{\it Single differential charged current DIS cross section
 as a function of $Q^2$. 
}
\label{fig:cc99}
\vspace{-1.0cm} 
\end{center}
\end{figure}
%

%% file: f2sum.tex
\section{Measurement of the Proton Structure Function}
\begin{figure}[htp] \unitlength 1mm
\vspace{1.5cm}
 \begin{center}
\begin{picture}(50,100)
 \put(-15,0){
 \epsfig{file=f2sum.ps,
 bbllx=56.,bblly=18.,bburx=534.,bbury=782.,clip,width=8.cm}
 }
%
 \end{picture}
\vspace{-1.1cm}
 \caption{\it The proton structure function $F_2$ as a function
 of $Q^2$ for bins in the Bjorken
 scaling variable $x$. Superimposed
 is a NLO QCD fit.}
\label{fig:f2sum}
 \end{center}
\vspace{-1.0cm}
\end{figure} 
A classic key measurement for the understanding of the proton
structure is the precise determination of $F_2$ by counting
inclusively the lepton scattered off the proton.
The kinematic domain spans about $5$ orders of magnitude
in \Qsq from $ 1 \lsim \Qsq \lsim 30000 \GeVsq$
and covers the momentum range from the sea quark region
at low $x$ ($ x \gsim 2 \cdot 10^{-5}$) 
to the valence quark region at large $x$
($x \lsim 0.65$).

An inclusive measurement has the advantage that it
can be directly compared to QCD calculations.
However, small interesting effects on top of the
dominating parton evolution are difficult to reveal.
Therefore $F_2$ has to be precisely pined down to get
a handle on the QCD evolution and to constrain
the non-perturbative parton density functions.
A summary of the presently available data is
shown in Fig.~\ref{fig:f2sum}.
For $x \le 0.01$ the statistical error is about 
$\lsim 1\%$ and the systematic accuracy has reached 
$\lsim 3-4 \%$ in most of the covered range.
The largest experimental systematic errors are
introduced by the hadronic and electromagnetic energy scale
uncertainties.
Compared to previous H1 results~\cite{h1f296,h1f297a,h1f297b}
the systematic uncertainties improved by nearly a factor 
of two. This precision can compete with the
fixed target data.
 
The measurement extents to energies of the scattered
positron down to $3$ \GeVx. This was only possible
by using the
backward electromagnetic fibre-lead calorimeter 
SPACAL (operated since $1995$) and
by using the backward silicon strip detector 
(operated since $1997$)
to measure tracks within $172^o < \theta_{el} < 177^o$.
It gives access to the kinematic region of low \Qsq
and high $y \lsim 0.75$ where the DIS cross section
is sensitive to $F_L$.
A new method has been introduced 
to improve the sensitivity to $F_L$ by measuring
the derivative: 
$$
\frac{\partial \sigma}{\partial \log{y}} = 
\frac{\partial F_2   }{\partial \log{x}} - F_L \cdot
\frac{2 - y        }{Y_+^2} + 
\frac{\partial F_L }{\partial \log{x}} \cdot
\frac{y^2}{Y_+}
$$
at fixed \Qsqx. Assuming that $\partial \sigma/\partial \log{y}$
is a linear function of $\log{y}$ up to large $y$,
the contribution of $F_2$ at high $y$ has been extrapolated
by a straight line fit for $y < 0.2$. The extracted $F_L$ 
extents down to
$\Qsq = 3 $ \GeV at $x \approx 10^{-4}$
and is in good agreement with previous H1 results~\cite{h1:fl96}
and consistent with $F_L$ as calculated by NLO QCD.

The NLO fit shown in Fig.~\ref{fig:f2sum} is performed in 
the $\overline{MS}$ renormalisation
scheme using the NLO DGLAP evolution equations
for three light flavours with the charm contribution
added in the massive scheme according to a NLO calculation
of the boson-gluon fusion process. 
The input distributions of the valence ($u_v, d_v$) 
and sea ($S= \bar{u} = \bar{d} = 2 \bar{s}$)
quarks and the gluon are parameterised at a
starting scale $Q^2_0 = 2 \GeVsq$ using $5$ parameters
which are determined in a fit to the data.
The muon-proton
data of BCDMS~\cite{bcdms} and the muon-deuteron
data of NMC~\cite{nmc} are included in the fit
to constrain the high $x$ behaviour of the parton
distributions. Target mass corrections were
applied for the fixed target data.
The strong coupling constant was
set to $\alpha_s(M_Z^2) = 0.118$ and the charm
quark mass to $m_c = 1.5$ \GeVx.
This fit gives a good description of the data for
all \Qsq and $x$. At low $x$ the behaviour
of $F_2$ is dominated by the gluon and the sea
quark distributions. No need for a parton evolution
different from NLO DGLAP is found.

%% file: jets.tex
\section{Jet production in DIS}
\begin{figure}[htp] \unitlength 1mm
 \begin{center}
 \epsfig{file=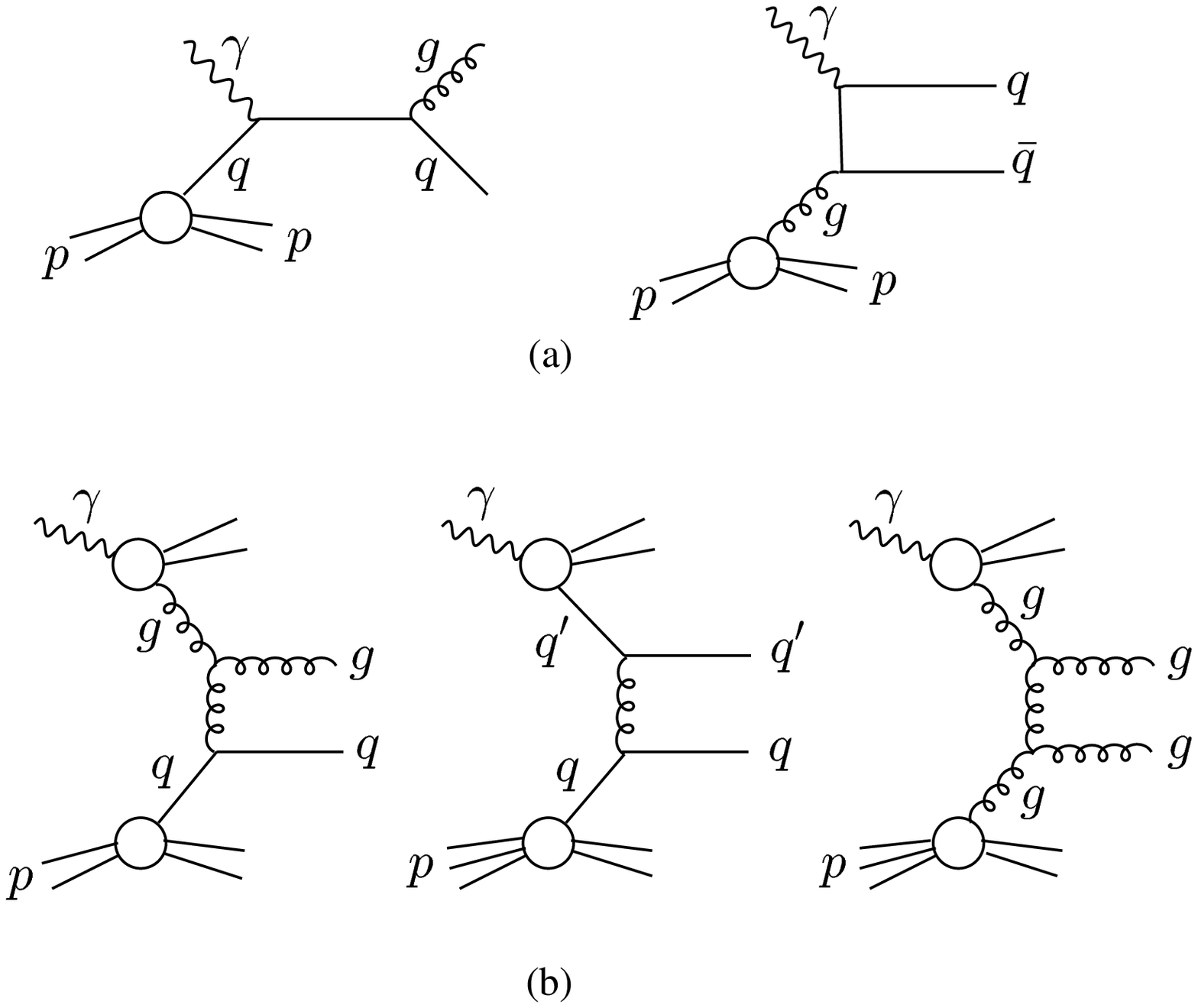,
 bbllx=44,bblly=468,bburx=484,bbury=599,clip,width=8.cm}
 \end{center}
\vspace{-1.0cm}
 \caption{\it Leading order feynman diagram for dijet
 production in DIS at HERA.}
\label{fig:feyjet}
\vspace{-0.9cm}
\end{figure}

The high centre of mass energy of HERA leads to a
large phase space for hadron production and to 
the possibility to observe clear jet structures
in DIS. Jets are infrared and collinear safe observables
defined by an algorithm
relating the unobservables quarks and gluons to the 
sprays of hadrons observed in the detector.
Events with two jets in the central part of the
detector can be produced in quark ($q \gamma \to q g$)
or gluon ($g \gamma \to q \bar{q}$) induced 
hard processes.
The leading order Feynman diagrams for dijet production 
are shown in Fig.~\ref{fig:feyjet}.
Dijet cross sections are directly sensitive to the strong
coupling constant $\alpha_s$ and to the gluon densities
in the proton. They can be calculated in pQCD to NLO
in $\alpha_s$ using numerical methods implemented
in several Monte Carlo programs. 
In a frame\footnote{e.g. the hadronic centre of mass
frame defined by $\vec{\gamma} + \vec{p} = 0$ or the
Breit frame defined by $\vec{\gamma} + 2 \; x \; \vec{p} = 0$.},
where the virtual photon $\gamma$ and the proton $p$ collide
head-on, it is obvious that two scales can characterise
the hardness of the process: $Q^2$
as used in the QCD analysis of the
inclusive DIS cross section or the transverse 
momenta of the two jets $E_T$ as used in hadron-hadron 
collisions.

H1 has reported the first fundamental understanding
of jet production in DIS at HERA energies
at the DIS 98 conference by demonstrating
that dijet cross sections for 
$ 10 < \Qsq < 5000 \GeVsq$ and 
$8.5 < E_T < 35$~\GeV
can be described by NLO QCD. 
Jets were defined by the inclusive $K_T$ 
algorithm~\cite{jet:inclkt,jet:cadosewe93} 
in the Breit frame
requiring two jets with $E_T > 5$ \GeV
and $E_{T,1} + E_{T,2}> 17$~\GeVx. 
Only jets lying well within the detector acceptance
$ -1 < \eta_{\sf jet,lab} < 2.5$ are considered.

At this conference further progress has been reported.
The double differential
single inclusive jet cross section $d\sigma/dQ^2 dE_T$
for $7 < E_T < 50$ \GeV 
has been made available. It is observed that
for higher \Qsq the $E_T$ spectrum gets harder.
Moreover dijet rates $1/N dn/dy_2$, where $y_2$ is the
maximal value of the resolution parameter which allows
to resolve two jets in an event, 
have been measured 
for $ 150 < \Qsq < 5000 \GeVsq$ using the 
Durham $K_T$ algorithm~\cite{jet:ktalogodis}  
with a scale of $100$ \GeVsqx.
Also the $y_2$ spectrum gets harder as \Qsq increases, i.e.
harder jet structures are resolved.
Both observables are well described by NLO QCD.
Furthermore it has been demonstrated that the 
dependences of the internal jet structure on the
$E_T$ and $\eta$ is well reproduced by 
QCD models~\cite{jet:h1jetstr}.  

\begin{figure}[htb] \unitlength 1mm
 \begin{center}
\mbox{\hspace{-0.5cm}{
 \epsfig{file=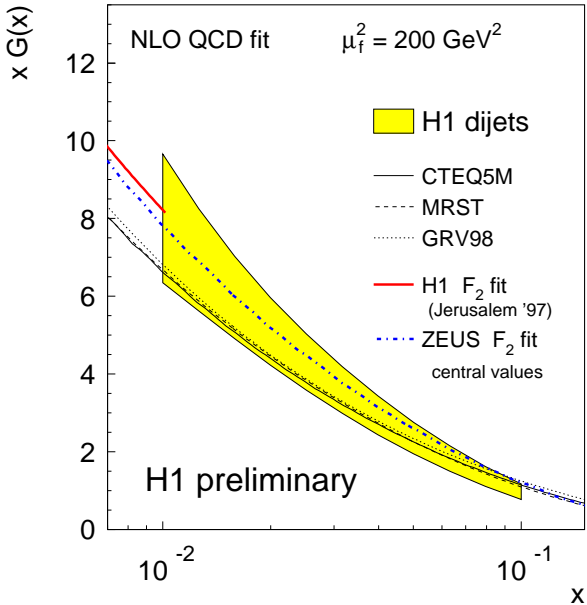,width=8.cm}
}}
 \end{center}
\vspace{-1.5cm}
\caption{\it The gluon density in the proton 
as a function of $x$ determined in a NLO pQCD
fit to dijet cross sections.
}
\label{fig:jetgluon}
\vspace{-1.0cm}
\end{figure} 
%
The dijet cross sections $d^2 \sigma/d\Qsq d\xi$
and $d^2 \sigma/d\Qsq dx$ have been used to extract
the gluon density in the proton
at a factorisation scale of $\mu_f = 200$ \GeVsq
which corresponds roughly to $\av{E_T^2}$.
$\xi$ is defined as 
$\xi  = x \; ( 1 + \hat{s}/Q^2)$, where
$\sqrt{\hat{s}}$ is the invariant mass of the dijet system.
For $\Qsq > 200$ \GeVsq $55\%$ of the dijet cross
section is caused by gluon induced processes.
In the fit $\alpha_s(M_Z)$ is assumed to be
$\alpha_s(M_Z) = 0.119 \pm 0.005$~\cite{jet:catani97a}.
This value has been mostly determined in processes which are 
independent of the proton structure. 
Inclusive DIS cross section at $200 < \Qsq < 650$ \GeVsq
are simultaneously fitted. These data strongly
constrain the quark densities, but
depend only weakly on the gluon density.  
The resulting gluon density together with its
error band is shown in Fig.~\ref{fig:jetgluon}
in the range $0.01 < x < 0.1$.
The error band is dominated by the uncertainty 
on $\alpha_s$, by the renormalisation scale dependence
and by the experimental absolute hadronic energy scale.
Good agreement with indirect determinations via scaling
violations of $F_2$ is found, but larger $x$-values
are reached. Gluon determinations from global
fits~\cite{th:mrst,th:cteq5,th:grv98} tend to give
slightly lower values, but are consistent within the
errors.

The precise knowledge of the gluon density at large
$x$ and of $\alpha_s$
is an important ingredient to predict cross
sections at the LHC~\cite{stirlinglhc}. 
The region $10^{-3} < x < 10^{-1}$ is directly
relevant for a Higgs with mass $100-500$ \GeV produced
at $| \eta | < 2$. 
For instance, the cross section for
the process $g g \to {\rm Higgs}$ gives within $20$\%
different results when different PDF parameterisations
are used. \Qsq has to be extrapolated by DGLAP by $3$ 
orders of magnitudes from the HERA region 
to reach values relevant at LHC.
Therefore the precise knowledge of the parton
evolution is also a key ingredient in the cross section
prediction.
Also the cross section $ g g \to W$ which can
be used as a luminosity monitor in proton-proton
collisions varies within $5\%$
when the most recent PDF of MRS and CTEQ are used.
However,
it is clear that comparing different PDF parameterisation
is rather inadequate to estimate their uncertainty,
since similar data are used in the fit and similar assumptions
are made. Therefore, it is also important to develop
techniques to extract QCD parameters together with
an error estimate. Within one experiment H1 has 
demonstrated that this is perfectly possible and more 
results are expected in the future.

%% file: gluonpho.tex
\section{The Structure of the Real Photon}
\begin{figure}[htp] 
\vspace{-1.5cm}
 \begin{center}
 \epsfig{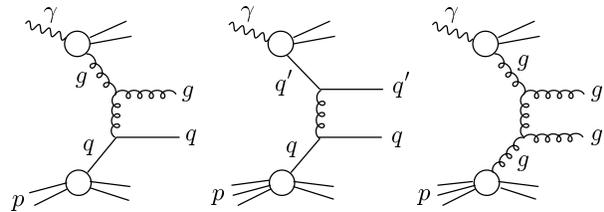}
 \end{center}
\vspace{-1.5cm}
 \caption{\it Feynman diagram for resolved processes 
 in dijet production  at HERA.}
\label{fig:feyjetres}
\vspace{-0.8cm}
\end{figure} 
%
In the quark parton model DIS is viewed as a highly virtual photon 
interacting with partons freely moving in the proton.
This is a good approximation when small distances are 
probed, i.e. in the limit where \Qsq is large.
In this regime,
the photon behaves like a point-like object, i.e. it 
directly couples to quarks to produce the hard scattering.
At low virtualities the photon dominantly 
fluctuates into vector mesons. However, 
the photon may also fluctuate into a
$q \bar{q}$ state with higher transverse energy without 
forming a bound state. In this case the photon acts as a 
source of strongly interacting partons. 
Such a 'resolved' process can be calculated within pQCD using the concept 
of a photon structure function.

They can be experimentally investigated by demanding
jets or charged particles at high transverse energy in 
the final state. Such data are sensitiv to the quark and
to the gluon content in the photon. Of particular interest
is the behaviour of the photon at low fractional
energies of the parton participating in the
hard scattering ($x_\gamma$). 
\begin{figure}[htp] 
\vspace{4.2cm}
 \begin{center}
\mbox{\hspace{-0.5cm}
 \epsfig{file=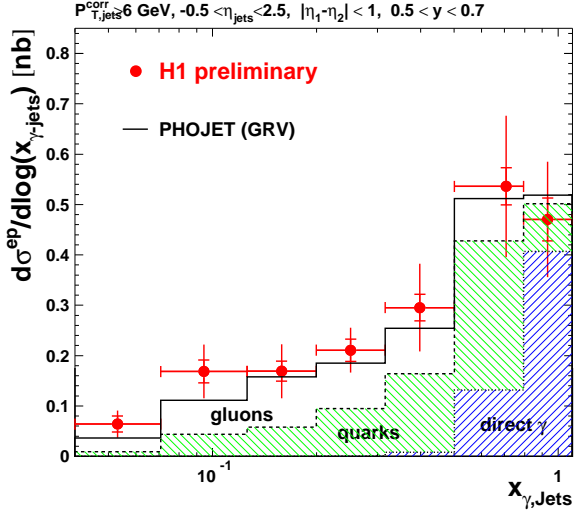,
 bbllx=8,bblly=164,bburx=534,bbury=305,clip,width=8.cm}}
 \end{center}
\vspace{-1.5cm}
 \caption{\it Dijet cross section as a function of 
 $x_{\gamma}$.}
\label{fig:dsigmadxgamma}
\vspace{-0.8cm}
\end{figure} 

The access to low $x_{\gamma}$ 
is experimenally difficult. Since $x_{\gamma}$ is given by 
$ x_{\gamma} 
= 
\sum_{1,2} E_T \; e^{-\eta} 
/
2 \; E_\gamma 
$, 
low jet $E_T$ destroying a good correlation with the hard partons 
or large $\eta$ values limited by the detector acceptance
and the understanding of soft underlying processes
have to be used. 
In a recent analysis of photoproduced events
($\Qsq \approx 0 \GeVsqx$) jets are selected using a
cone algorithm~\cite{jet:cdf_92} with $R=0.7$.
After correction for energy steming from soft processes
migrating into the jet cone (``pedestal'')  
$E_T > 6$ \GeV is required. Both jets have
to lie well within the detector acceptance of
$-0.5 < \eta_{jet}< 2.5$ and their rapidity difference
must not exceed $|\Delta \eta | < 1$. 
The dijet cross section is shown as a function of 
$x_{\gamma}$ in Fig.~\ref{fig:dsigmadxgamma}.
It is compared to the LO prediction of the 
PHOJET~\cite{mc:phojet95,mc:phojet96}
QCD model using the GRV-LO PDF~\cite{th:grv92}
for the photon and the proton.
While at $x_{\gamma} \approx 1$ the direct contribution
dominates, for $x_{\gamma} < 1$ the resolved contribution
is most important. At the lowest $x_{\gamma}$ values 
gluon initiated processes are mainly responsible for the measured
cross section.

\begin{figure}[h] \unitlength 1mm
\vspace{-1.1cm}
 \begin{center}
 \epsfig{file=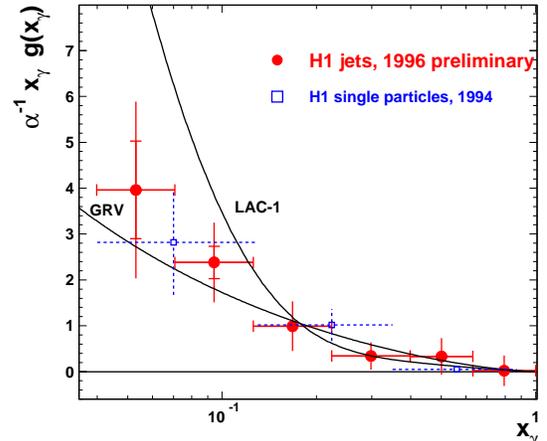,width=7.cm}
 \end{center}
\vspace{-1.5cm}
 \caption{\it Extracted leading order gluon density in the photon
 as a function of $x_{\gamma}$.}
\label{fig:gluonxgamma}
\vspace{-1.0cm}
\end{figure}

A leading order interpretation of the data can be made
by using the concept of an effective PDF for which the
dijet cross section can be written as~\cite{maxwell84}:
$$
\vspace{-0.3cm}
\frac{f^{\gamma/e}(y,Q^2)}{y} 
\cdot \sum_{i,j}^{N_f}
\frac{f_{\rm eff}^{i/\gamma}(x_\gamma,E_T^2,Q^2) 
}{x_\gamma} 
\cdot 
\frac{f_{\rm eff}^{j/p}}{x}  
\cdot 
|ME_{ij}|^2 
$$  where the sum runs over all scattering matrix
elements depending on the scattering $\cos{\theta^*}$
of the processes shown in Fig.~\ref{fig:dsigmadxgamma}
folded by the density of parton
$j$ in the proton and parton $i$ in the photon. 
The effective probability to find parton $i$ in the photon
can be written as:
$$
f^{i/\gamma}_{eff} = \sum^{N_f} (q + \bar{q}) + \frac{9}{4} g. 
\vspace{-0.3cm}
$$
The unfolded effective gluon distribution is shown in 
Fig.~\ref{fig:gluonxgamma} as a function of $x_\gamma$.
The gluon density rises as $x_\gamma$ decreases. This
rise is less pronounced than predicted by the
LAC1 structure function~\cite{th:lac1},
but is in agreement with the GRV expectation~\cite{th:grv92}.
The results improve a previous measurement~\cite{h1:gluonphoton95}.
and agree with earlier results using
the complementary approach of
single charged particles~\cite{h1:chargedpartgluon}.
%

%% file: eventsha.tex
\section{Event Shapes in Current Region of Breit Frame } 
Measurements of event shape variables $F$ provide information
about perturbative and non-perturbative aspects of QCD.
They allow to fit analytical expressions to data
without referring to a fragmentation model by
exploiting their characteristic power behaviour. 
Event shapes have been extensively studied in 
\ee experiments at different center of mass 
energies~\cite{jet:bethke97}.
%
%
%
%
\begin{figure}[htb] 
\vspace{-0.1cm}
  \epsfig{file=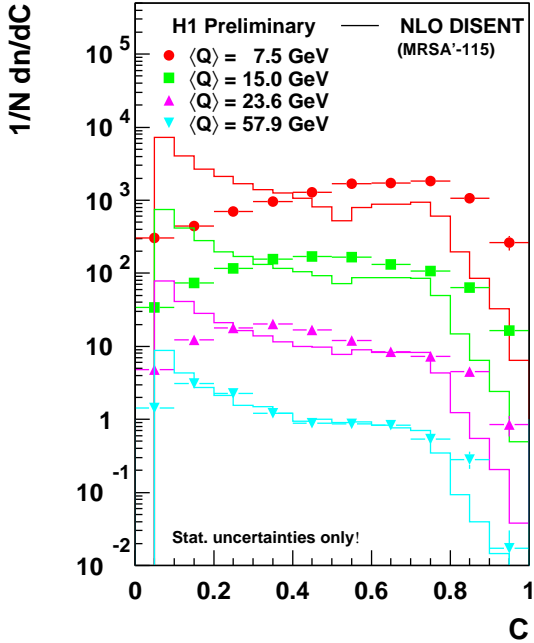,width=7.5cm}
\vspace{-1.5cm}
\caption{\it Distribution of the $C$-parameter for various
 $Q = \sqrt{Q^2}$ bins.}
\label{fig:cpara}
\vspace{-0.9cm}
\end{figure}

\begin{figure}[htb] 
\vspace{-0.7cm}
  \epsfig{file=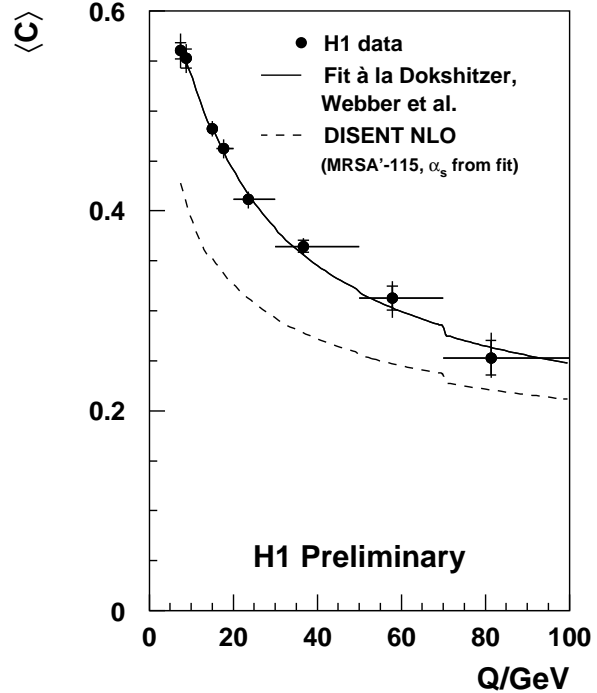,width=8.cm}
\vspace{-1.5cm}
\caption{\it Mean value of the $C$-parameter as a function
of $Q=\sqrt{Q^2}$.}
\label{fig:cparamean}
\vspace{-1.0cm}
\end{figure}

Results from \ee can be compared to DIS in
the current hemisphere of the Breit frame.
Thanks to the large kinematic range covered at HERA,
the dependence of mean event shape values 
on a hard scale can be studied in one experiment.
Event shapes like thrust, jet broadening and
the jet mass have been investigated in the 
past~\cite{h1:eventshapes}. In this workshop
for the first time also the $C$-parameter and
differential jet rates $y_2$ for the JADE and for the
$K_T$ jet algorithm have been presented.
The $C$-parameter is defined as:
$
C :=
3 ( \lambda_1 \lambda_2 + \lambda_2 \lambda_3 + \lambda_3 \lambda_1)
$ 
where the $\lambda_i$ are the eigen values of the 
linearized momentum tensor of the final state particles.
As an example, the distribution of the 
$C$-parameter in bins of the
momentum transfer $Q = \sqrt{Q^2}$ ranging 
from $7$ to $100$ \GeV is shown in Fig.~\ref{fig:cpara}.
While at high $Q$ the data
are described by a ${\cal O}(\alpha_s^2)$ 
calculation~\cite{mc:disent97}, at low $Q$
the shape does not agree with perturbative QCD.
The same conclusions hold for
the mean value of the $C$-parameter shown as a function 
of $Q$ in Fig.~\ref{fig:cparamean}.

The mean event shapes can be expressed by a 
perturbative and a non-perturbative contribution
of the form~\cite{th:powercorr,th:powerdis}:
\vspace{-0.2cm}
\begin{eqnarray}
 \fmean = \fpert + \fpow\, \hspace{3.1cm} \nonumber \\
  \fpert = c_{1,F}(x) \, \as(Q) + c_{2,F}(x) \, \as^2(Q)
\hspace{0.6cm} \nonumber \\
\fpow = a_F \frac{16}{3 \pi} \frac{\mu_i}{Q}
\ln^p{\frac{Q}{Q_0}} \;
\left [ \overline{\alpha}_0(\mu_i) - \alpha_s(Q) - \nonumber 
\right. \\  \left.  
\right ( b \ln{\frac{Q^2}{\mu_i^2}} + k + 2b \left ) \; \alpha_s^2(Q)
\right ] \nonumber
\label{eq:fpow}
\vspace{-0.1cm}
\end{eqnarray}
where $b=(11\,C_A-2 f)/12 \pi$ and 
$k=\left [ (67-3 \pi^2) \, C_A-10 \, f \right ]/36 \pi$
and
$\overline{\alpha}_0$ is a free, but 'universal', effective  coupling
parameter below an 'infra-red' matching scale 
$\Lambda_{QCD} \ll \mu_i \ll Q$.

Such an ansatz is able to describe the data
(see Fig.~\ref{fig:cparamean}).
The power corrections are large at low $Q$, but become less 
important with increasing $Q$. The parameters
$\alpha_s$ and $\overline{\alpha}_0$
can be simultaneously fitted to the data. It turns out that
the analytical form of the power correction is adequate 
to describe the data for all studied event shapes.
A simple form $ \fpow \sim 1/Q$ is not able to describe the data.

\begin{figure}[htb] 
 \begin{center}
  \epsfig{file=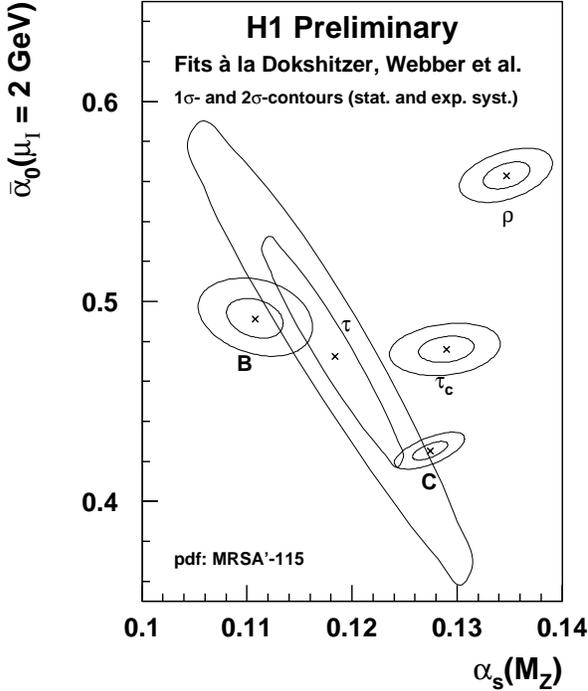,width=8.cm}
 \end{center}
\vspace{-1.3cm}
\caption{\it $\alpha_s(M_Z)$ versus $\overline{\alpha}_{0}(\mi)$
from the fit of four different event shapes variables. 
The ellipse gives the $1 \sigma$ and
$2 \sigma$ contour of the 
statistical and the systematical error.}
\label{fig:ellipse}
\vspace{-1.cm}
\end{figure}
The question whether consistent fit results
are obtained when using different event shape
variables is addressed in Fig.~\ref{fig:ellipse}.
The ellipse displays the $1 \sigma$ and $2 \sigma$ 
contours of the statistical and the systematical error.
The NLO dependence on the renormalisation scale
is not included. The experimental errors are highly 
correlated.
The two parameters 
$\overline{\alpha}_{0}(\mi)$ and $\alpha_s(M_Z)$
come out about in the same range like found in \ee
collisions. However, in particular the
large spread of $\alpha_s$ is not fully satisfactory.
More precise data and more theoretical work is
needed to continue this promising way to get an (analytical) 
understanding of hadronisation for specific variables.
Open questions are the correct inclusion of higher orders
to get a more consistent picture and the
treatment of the uncertainty introduced by the parton
density function to the fit result.

%% file: pi0.tex
\section{QCD parton dynamics at low $x$ } 
\begin{figure}[htp] \unitlength 1mm
 \begin{center}
  \begin{picture}(80,30)
   \put(-10,0){
    \begin{tabular}{cc}
      \epsfig{figure=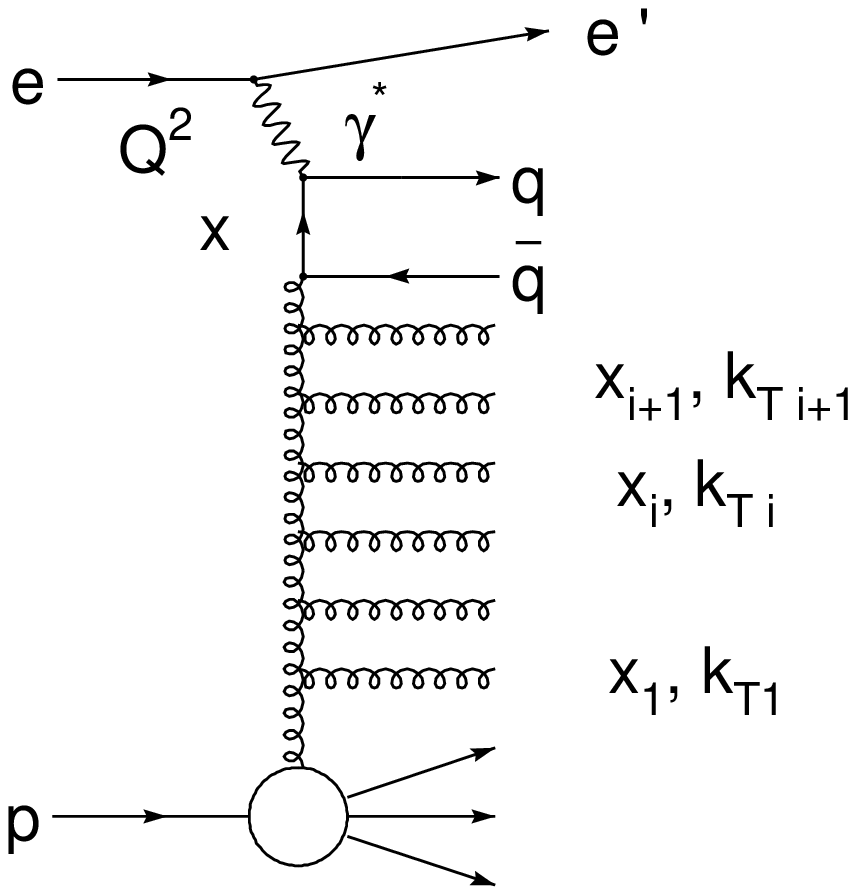,width=6cm}
      \mbox{\hspace{-4cm}
      \epsfig{figure=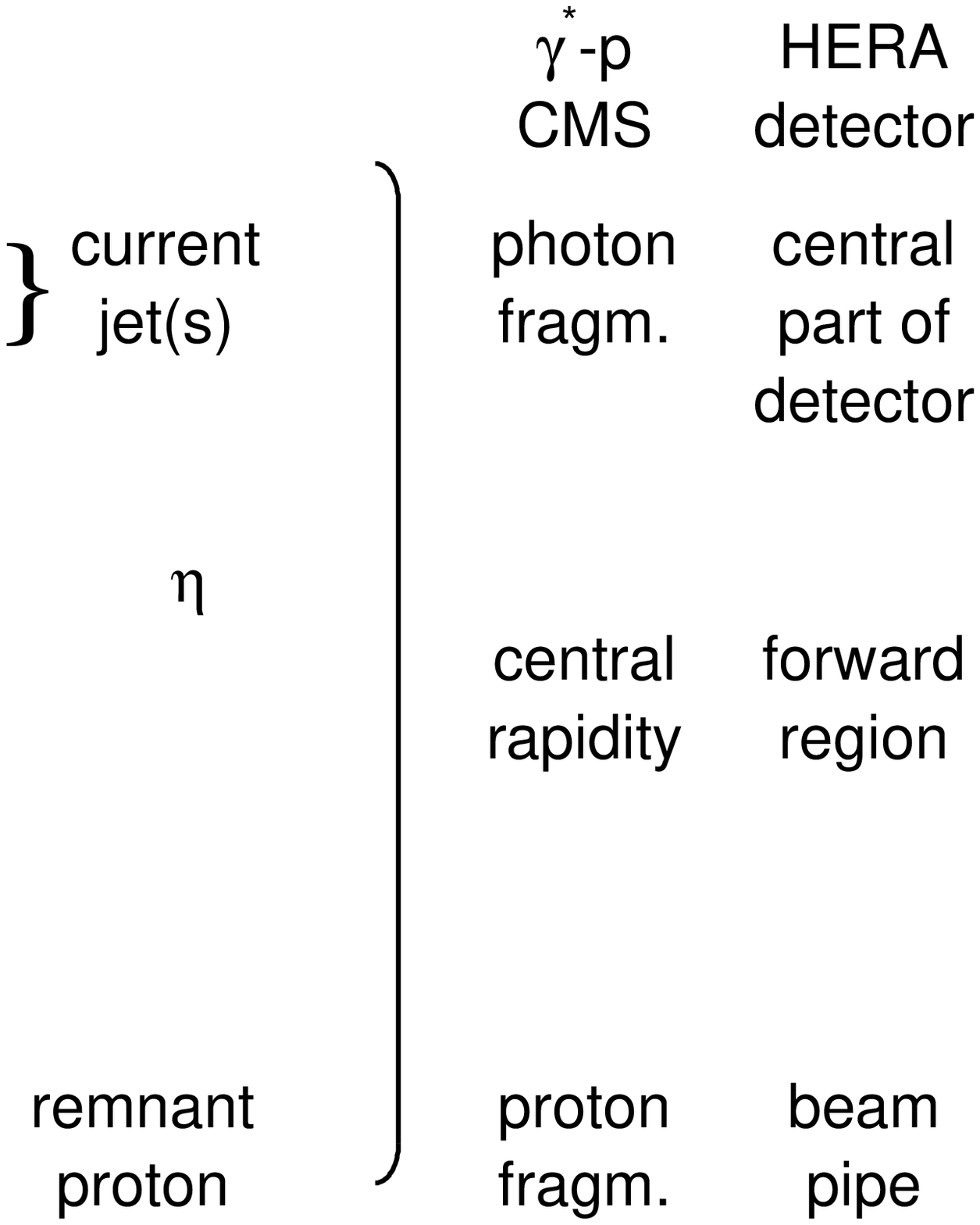,width=6cm}
      }
    \end{tabular}
   }
  \end{picture}
 \end{center}
\vspace{1.0cm}
\caption{\it Diagram of a DIS events at low $x$.}
\label{fig:gluonladder}
\vspace{-0.8cm}
\end{figure}
%
 At low $x$ the simple picture of DIS as a process where
 a virtual photon interacts instantaneously with a point-like
 parton freely moving in the proton has to be modified.
 The phase space for gluon emission ($W^2 \approx Q^2/x$) 
 between the photon and the proton becomes so large that
 many partons can be radiated before interacting with the 
 photon. Such an interaction is illustrated in 
 Fig.~\ref{fig:gluonladder}. 
 In particular in the central
 rapidity region the detection of hard particles can 
 discriminate between different evolution 
 schemes of the parton cascade~\cite{th:kuhlenlowx}. 
 
 The transitions $g \to gg$ and $g\to q\bar{q}$ at each point
 in the ladder can be approximated by the DGLAP 
 equations~\cite{th:dglap}. 
 By resumming the ${(\alpha_s \ln{Q^2})}^n$ terms 
 they predict the \Qsq evolution of a parton known to be 
 point-like at some given scale $Q^2_0$ to the region where 
 the interaction with the photon takes place. 
 To derive the DGLAP equations a strongly ordered
 configuration in the parton virtualities along the ladder
 has to be assumed. This leads to a suppression of the available
 phase space for gluon radiation towards the proton.
 
 When the ${(\alpha_s \ln{1/x})}^n$ terms become large, they
 have to be taken into account, e.g. by the resummation
 accomplished by the BFKL equations~\cite{th:bfkl}.
 In a physical gauge, these terms correspond
 to an $n$-rung ladder diagram in which gluon emissions
 are ordered in longitudinal momentum. The strong ordering of
 the transverse momenta is replaced by a diffusion pattern
 as one proceeds along the gluon chain.
 The BFKL equations describe how a
 parton in the proton is dressed by
 a cloud of gluons at low \xB~localised in a fixed transverse 
 spatial region of the proton. 

%
\begin{figure}[htb]
\vspace{-0.2cm}
\begin{center}
 \epsfig{file=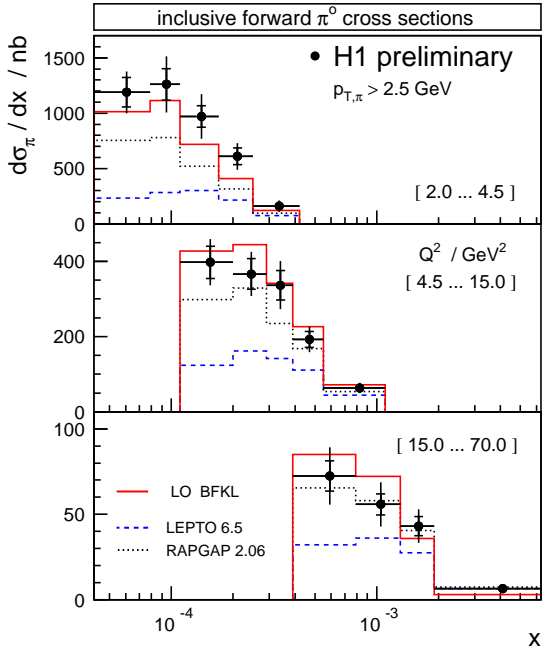,width=7.5cm}
\end{center}
\vspace{-1.4cm}
\caption{\it $\pi^0$ cross section as a function of $x$.}
\label{fig:inclpi0}
\vspace{-0.6cm}
\end{figure}
The density of
hard partons in the central rapidity region can be
experimentally explored by identified particles or jets 
in the forward region of the H1 detector 
(see Fig.~\ref{fig:gluonladder}). 
The finely granulated H1 LAr calorimeter with its
$3.5 \times 3.5$\cm calorimeter cells and a $4$ fold 
segmentation offers the opportunity to identify high 
energetic $\pi^0$ mesons by exploiting the typical
properties of electromagnetic showers.
The high particle density in the forward region 
for $5^o < \theta_{lab} < 25^o$ 
requires a good understanding of the detector and is the
main experimental challenge.
The inclusive cross section as a function of $x$ 
for $\pi^0$ mesons with a $\pt > 2.5$ \GeV and
$x_\pi = E_\pi/E_p > 0.01$ in the centre
of mass system for three \Qsq regions ranging from 
$2.0 < \Qsq < 70$ \GeVsq is shown in Fig.~\ref{fig:inclpi0}.
The cross section rises by a factor $8$ towards
low $\xB$ for each \Qsqx. In the lowest $\Qsq$ region 
values down to $x = 5 \cdot 10^{-5}$ are reached.
A weak dependence of the cross section on \Qsq
is found and for each \Qsq the $x$ spectrum is
similar.
A Monte Carlo study indicates that a 
$\pi^0$ with $\pt \approx 2$ \GeV originates on average
from a parton with $ \pt \approx 5$ \GeVx.

The LEPTO Monte Carlo~\cite{mc:lepto} model based on the 
leading order matrix element (${\cal O}(\alpha_s)$)
$e q \to e q g$ or $e g \to e q \bar{q}$ 
incorporating higher order emissions from the proton side
in a leading logarithm 
approximation close to the DGLAP equations is not able
to describe the data. In the highest \Qsq bin
LEPTO is below the data by a factor of $2$. 
This difference increases towards lower \Qsqx.
The RAPGAP model~\cite{mc:lepto} 
includes in addition LO resolved photon 
processes. It is in much better agreement with the data.
However, at the lowest $x$ and \Qsq also this model fails
to describe the data. An analytical calculation based on the
leading order BFKL equation~\cite{th:bfklpi0}  
agrees with the data. In this calculation the parton
calculation is related to the measurement using a 
fragmentation function.

Whether this measurement can be taken as a proof that
BFKL effects are needed to describe HERA data remains
an open question. It is not fully transparent what
approximations have been included in the calculations
in view of the large NLO correction to the BFKL 
equation.
Moreover it is not a priori clear if the fragmentation
functions are valid in the high particle density environment
at HERA. Furthermore, it has been shown that very similar
measurements of jets in the forward 
region~\cite{h1:fwdjet,zeus:fwdjet} can be described
by a ${\cal O}(\alpha_s^2)$ calculation of direct and resolved 
processes~\cite{jet:potterfwdjet}. Since the photon
splitting term is the dominant contribution of the
resolved part, this calculation suggest that a
${\cal O}(\alpha_s^3)$  ´direct´ calculation would be enough to describe
the hadronic final state data at low $x$.
In this case the parton ladder is not long enough
such that resummation effects to all
order of the $\ln{(1/\xB)}$ terms are needed
in the HERA regime.

%
The strong rise of the inclusive $\pi^0$ cross section
reminds the rise of $F_2$ towards low $x$ 
which has been before the advent of HERA predicted
by the BFKL equations. 
We know today that $F_2$ can be well described by the 
NLO DGLAP
equations, but is also
compatible with BFKL.
It is interesting to note that the
ratio of the forward $\pi^0$ cross section to the
inclusive cross section is constant over the
full $x$ range. 
At $2 < \Qsq < 4.5$ \GeVsq approximately $0.2\%$
of the DIS events contain a forward $\pi^0$ with 
$\pt > 2.5$ \GeVx. At $15 < \Qsq < 70$ \GeVsq 
this ratio rises to $0.5 \%$.
While for an inclusive measurement the DGLAP evolution
is adequate, for observables where the parton evolution
is probed at a specific $\eta$ region the implementation
of the DGLAP equation in Monte Carlo models are too
restrictive. Only models allowing for a large phase
space of parton emission are in agreement with the data.

The constant (with respect to the inclusive cross section) 
probability to emit hard particles in the forward region
can be juxtaposed to the probability to emit
no particle at all. It is known since $1993$ that the
ratio of events exhibiting a large rapidity gap is
approximately $10 \%$ for $5 \le \Qsq \le 50$ \GeVsq
and $ 2 \cdot 10^{-4} \le \xB \le  2 \cdot 10^{-2}$.
Does this mean that there is a connection between the
production mechanism of rapidity gap events and
the parton dynamics at low \xB ?

%% file: diffchar.tex
\section{Diffraction}
A new era in understanding hadronic interactions
was opened by the observation that
a surprisingly large fraction of DIS events 
($\approx 10 \%$) 
exhibited a rapidity region free of hadronic activity
between the particles
emerged from the hard scattering and the proton.
Since in normal DIS large rapidity gaps are
exponentially suppressed, 
these events were soon analysed in terms of 
an exchange of a colourless object probed by
the virtual photon. If no quantum numbers are
exchanged such interactions are called ``diffractive''
and the colourless object is called ``pomeron'' ($\pom$)
\begin{figure} \centering
 \setlength{\unitlength}{1cm}
  \epsfig{bbllx=20,bblly=290,bburx=300,bbury=540,clip=,
            file=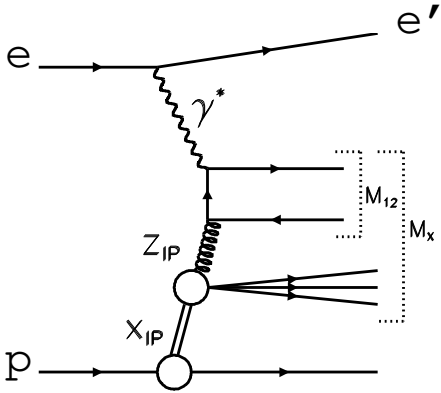,width=7.cm}
\vspace{-1.7cm}
\caption{\it Diagram for a colourless exchange in DIS.}
\label{fig:feyndiff}
\vspace{-0.5cm}
\end{figure}

The diagram for the production of diffractive
events is depicted in Fig.~\ref{fig:feyndiff}.
The hadronic final state can be split up in two
distinct systems $X$ and $Y$ which are separated by 
the largest rapidity gap in the event. 
Usually events are selected by requiring an
absence of activity in the forward part of the
detector such that
$Y$ has a small invariant mass
($M_Y \lsim 1.6 $ \GeVx) and 
the squared momentum transfer $t$ 
between the incoming proton and $Y$ is small 
($|t| \lsim 1 \GeVsqx$). 
The following kinematic variables can be
defined to describe the diffractive production
mechanism:
\begin{eqnarray}  
\beta  
\simeq \frac{Q^2}{Q^2 + \mx^2}, 
\; 
\xpom  
\simeq \frac{Q^2 + \mx^2}{Q^2 + W^2},
\;
\zpom \simeq \frac{Q^2+M_{12}^2}{Q^2+M_X^2}
\nonumber 
\end{eqnarray}  
$\xpom$ measures the fraction of the proton momentum
transfered to the $\pom$ and $\beta$ is
the momentum fraction of the $\pom$ momentum
carried by the quark coupling to the virtual photon.
If the system $M_X$ is further resolved in a subsystem
$M_{12}$, the variable
$\zpom$ gives the momentum fraction of the struck parton. 
If $M^2_{12} < M_X^2$, i.e. $\zpom < 1$, 
not the entire energy from the colourless exchange 
takes part in the hard scattering.

Usually the diffractive interaction is described in 
terms of Regge phenomenology and the kinematic
dependences are parameterised as:
\begin{eqnarray}  
\; \; \; \; \; \; \; \;
\frac{d\sigma}{dt dM_X^2} \propto 
{\left ( {W^2} \right )}^{2 \alphapom(t)-2} e^{b_0 t} 
\nonumber
\end{eqnarray}  
where $\alphapom(t)$ is the effective leading
Regge trajectory. For soft diffractive processes,
$\alphapom(t)$ takes the universal form
$\alphapom(t) \sim 1.08 + 0.2 \; t$.
When a hard scale is present, $\alphapom(0)$
($t$-slope) is expected to increase (decrease).
 
The ``soft'' pomeron ansatz successfully describes
the total and elastic hadron-hadron and
hadron-photon cross section as well as the
energy dependence of the cross section of light
vector mesons ($M_X=M_V$). 
However, a single soft pomeron does not describe all
diffractive data measured at HERA.
In presence of a hard scale like \Qsq or a larger vector
meson mass the cross section steeply increases with $\Wgp$.
A new type of dynamical pomeron may begin to play
a role whose structure can be tested in DIS.
HERA allows diffractive interactions to be analysed in 
terms of parton dynamics and 
offers the possibility to study the transition
between perturbatively calculable and incalculable
strong interactions.

\subsection{Inclusive Diffractive measurements}
The inclusive diffractive data are consistent with
a partonic interpretation of the pomeron for which
the cross section factorises into the parton density
of the pomeron and a pomeron flux factor
describing the probability to find a pomeron in the
proton~\cite{ingelschlein}. 
%
The parton densities can be determined by a QCD 
analysis~\cite{h1:h1pdfdiff}
of the inclusive diffractive cross section:
$$
\frac{d^3 \sigma_{ep \to e X Y}}{dQ^2 d\beta d\xpom}=
\frac{4 \pi \alpha^2}{\beta Q^4} ( 1- y + \frac{y^2}{2} )
F_2^{D(3)}(\xpom,Q^2,\beta).
$$
At fixed $\xpom$
the \Qsq dependence of the diffractive
structure function  $F_2^{D(3)}(\xpom,Q^2,\beta)$
is rather flat.
The $\beta$ dependence exhibits a logarithmic
rise characteristic for scaling violations.
When fixing the quark and gluon distribution at
a starting scale of $Q^2_0 = 3 \GeVsq$
and evolving them to larger \Qsq using the
DGLAP equations,
it is found that most of the partonic content of the 
pomeron is carried by hard gluons. 
At $\Qsq = 4.5$ \GeVsq the pomeron
consist to $90\%$ of gluon and even at
$\Qsq = 75$ \GeVsq the gluon fraction is still
$80\%$. Using these fits performed for $\Qsq < 100$ \GeVsq
also the new H1 data reaching \Qsq up to $800$ \GeVsq 
can be described.
However, in an analysis of very low \Qsq data
ranging down to 1 \GeVsq the logarithmic
$\beta$ dependence flattens off around
$\Qsq \approx 3-4$ \GeVsqx.
%
%
%
\subsection{Diffractive Dijet Production}
The partonic structure of diffractive interactions
can be further tested in the hadronic final state. 
High transverse momentum jet production 
is directly sensitive to the gluon content.
For $7.5 < \Qsq < 80$ \GeVsq and $0.1 < y < 0.7$
exactly two jets are selected with $E_{T,jet} > 5$ \GeVx.
Jets are defined by the cone algorithm.
$E_{T,jet}$ is measured  
relative to $\gamma^{\star}$ axis in rest frame of 
the system $X$.
%
\begin{figure}[h] 
\vspace{-1.3cm}
 \begin{center}
\epsfig{file=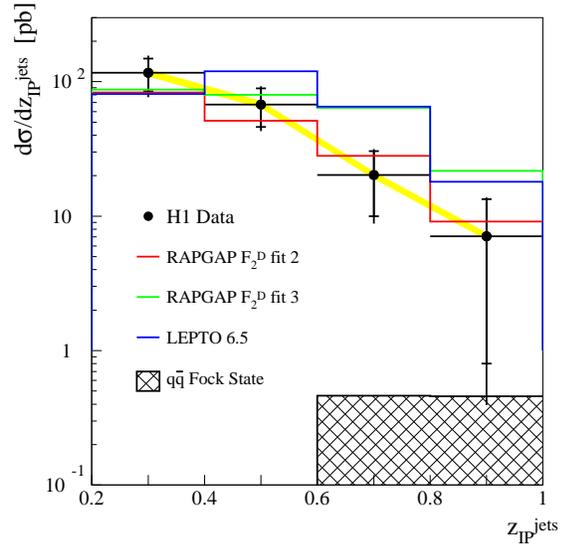,width=8.cm}
 \end{center}
\vspace{-1.2cm}
\caption{\it Diffractive jet cross section as a function of $\zpom$.}
\label{fig:diffjets}
\vspace{-1.0cm}
\end{figure}
The dijet cross section as a function of $\zpom$
is shown in Fig.~\ref{fig:diffjets}.
Models in which the pomeron is dominated by gluons
(fit 1 and 2 in ~\cite{h1:h1pdfdiff}) describe the data
well. Models where the pomeron consist only of
quarks at the starting scale undershoot the data by
a factor of $5$.

An alternative model sees the diffractive interaction 
as $q \bar{q}$ fluctuation of the photon
(decomposed in Fock states)
coupling via two gluons to the proton~\cite{bartelspom96,bartelspom96b} 
in a colour singlet
configuration ($2$-gluon model). 
At the largest $\zpom$ values,
where the full momentum of the $X$ system is
carried by the two jets and the model is expected
to be valid, it is somewhat
below the data, but still compatible with them
given the large systematic error.

In conclusion, dijet data support the gluon dominated
partonic picture of the pomeron.
\section{Diffractive Charm Production } 
\begin{figure} \centering
\vspace{-0.8cm}
 \setlength{\unitlength}{1cm}
 \epsfig{file=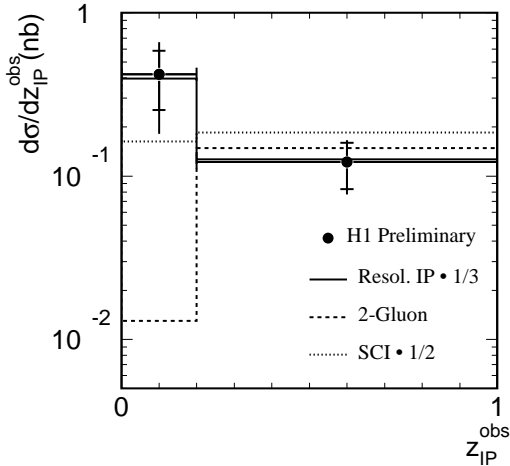,width=8.cm}
\vspace{-2.0cm}
\caption{\it Diffractive charm cross section as a function
of $\zpom$. The pomeron model has been divided by
a factor of $3$, the soft colour model by a factor
of $2$.}
\label{fig:diffcharmzpom}
\vspace{-0.9cm}
\end{figure}

The r\^ole of gluons in the diffractive production mechanism 
can be directly assessed by tagging charmed quarks
in the hadronic final state, since they predominately
originate from Boson-gluon processes.
Charm quarks are identified by reconstructing 
$D^{*\pm}$ mesons in the classical decay chain
$D^* \to D^0 \pi^+_{slow} \to (K^- \pi^+) \pi^+_{slow}$.
The branching ratio of this process is only $2.6\%$.
In the kinematical region of
$ 2 < \Qsq < 100 \GeVsq$ and $ 0.05 < y < 0.7 $ 
$45$ events have been selected 
in a data sample of about $21 \invpb$ taken in $1995-1997$.
$D^{*}$ mesons were required to have $p_T > 2$ \GeV
and $|\eta| < 1.5$.
The total cross section is measured to be:
$$
\sigma_{vis}(ep \to (D^* X)Y) = 
154 \pm 40 (stat) \pm 35 (syst) {\rm pb}.
$$
In Fig.~\ref{fig:diffcharmzpom}
the visible cross section
$\sigma_{vis}(ep \to (D^* X)Y)$ is shown as a function
of $\zpom$.
As for the diffractive jet data the cross section rises 
towards low $\zpom$. This means that in most of the events
a part of the hadronic final state produced in the
diffractive exchange is not associated with the hard
subprocess.

The shape of the $\zpom$ distribution is well reproduced
by the gluon dominated partonic pomeron picture.
This model remarkably reproduces the
rise of the cross section towards low $\zpom$.
However, the normalisation is off by a factor of $3$~!
Hence, much less charm quarks are produced in the
hadronic final state than predicted
by the parton density function extracted
from the inclusive diffractive measurements and applied
to the pomeron picture.

In the region where the hadronic system
$X$ dominantly consists of the $c \bar{c}$ pair alone
(i.e. $\zpom \approx 1$)
the $2$-gluon model gives a fair description of the 
data. 
As in the case of the diffractive dijets 
it is not able to describe the
the low $\zpom$ region. 
The soft colour interaction model where the colour
structure of a normal DIS events is modified by
a soft interaction leading to a colour singlet
in the final state~\cite{mc:scic} can not describe
the overall normalisation nor the shape of the $\zpom$
distribution.


In conclusion, the diffractive charm signal is 
more pronounced at low $\zpom$ values. The data
therefore indicate that not the entire colourless 
exchange couples to $c \bar{c}$ system. The behaviour
is well reproduced by the ``pomeron'' model, but the
overall measured cross section is three times lower
than expected. This is the first time that this picture
which has been so far supported by all inclusive and
hadronic final state data is put into question.  
More data are needed to clarify the situation.

%% file: vmmodel.tex
\section{Vector Mesons}
\begin{figure}[h] \unitlength 1mm
 \vspace{-1.2cm}
 \begin{center}
 \begin{picture}(0,100)
   \put(-40,95){{\large Elastic:}}
   \put(-45,54){\epsfig{file=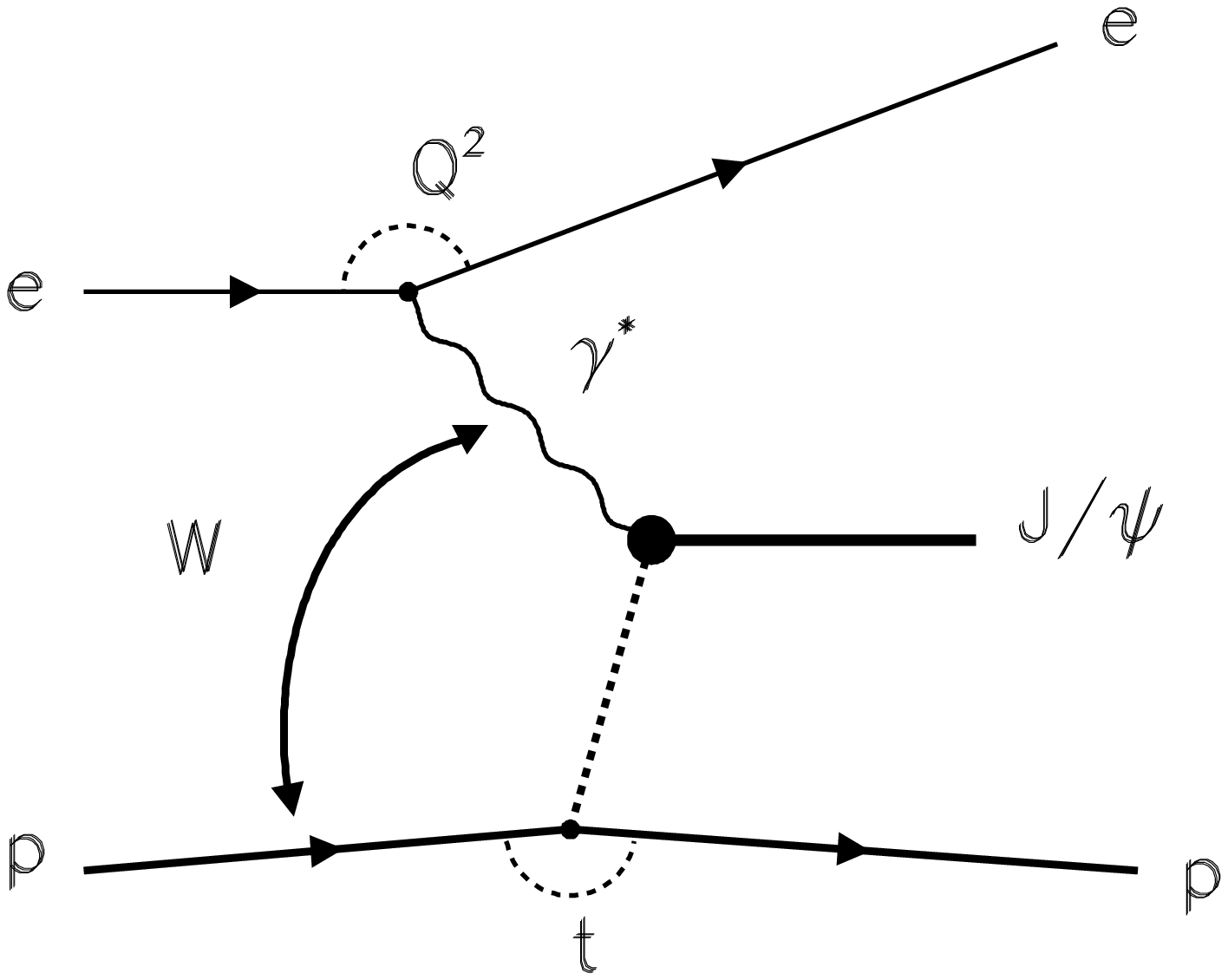,width=4.5cm}}
   \put(-5,54){\epsfig{file=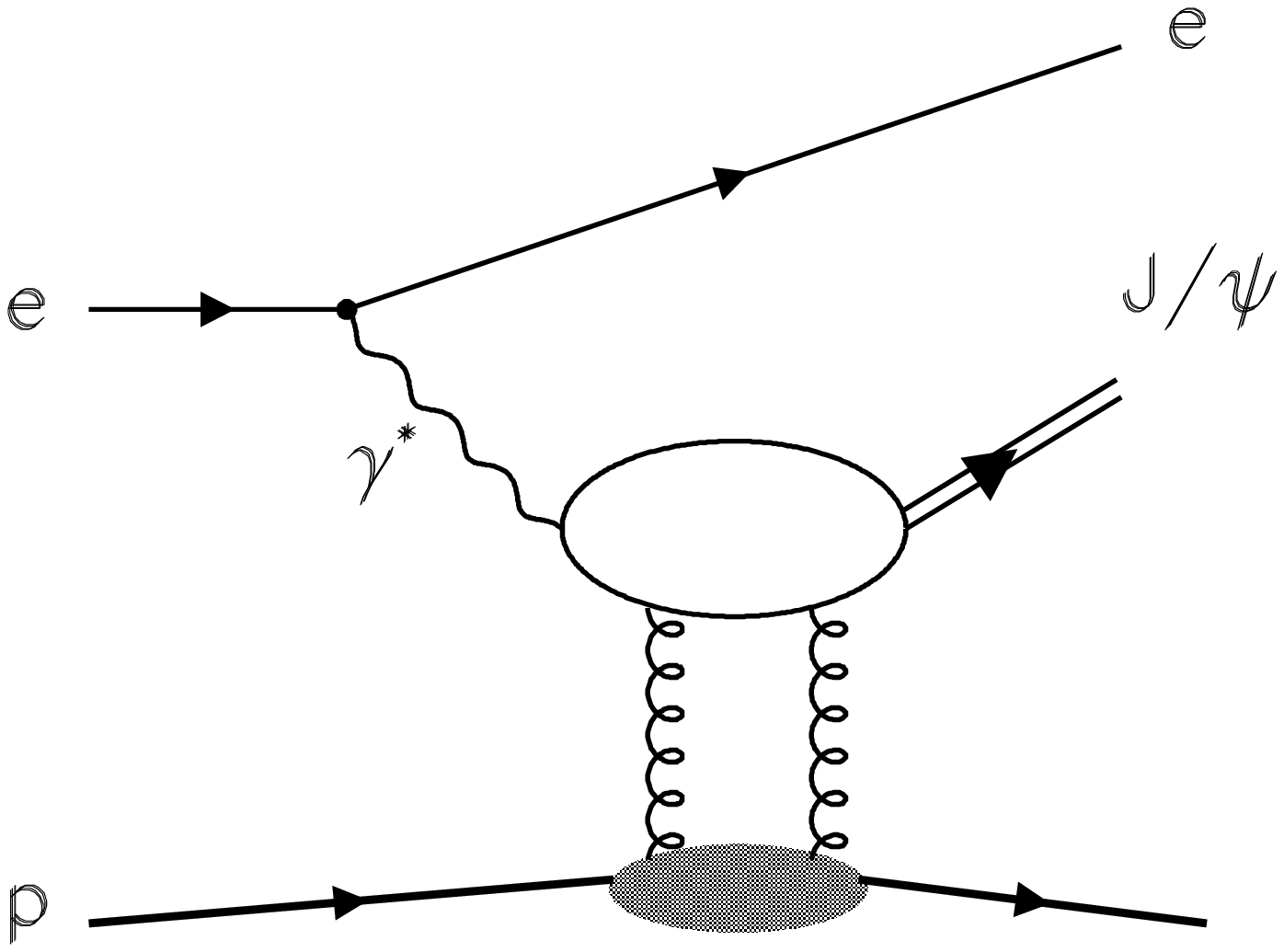,width=4.5cm}}
   \put(-40,60){{\large Inelastic: $X \ne$ proton}}
   \put(-30,25){Colour singlet}
   \put( 10,25){Colour Octet}
   \put(-45,15){\epsfig{file=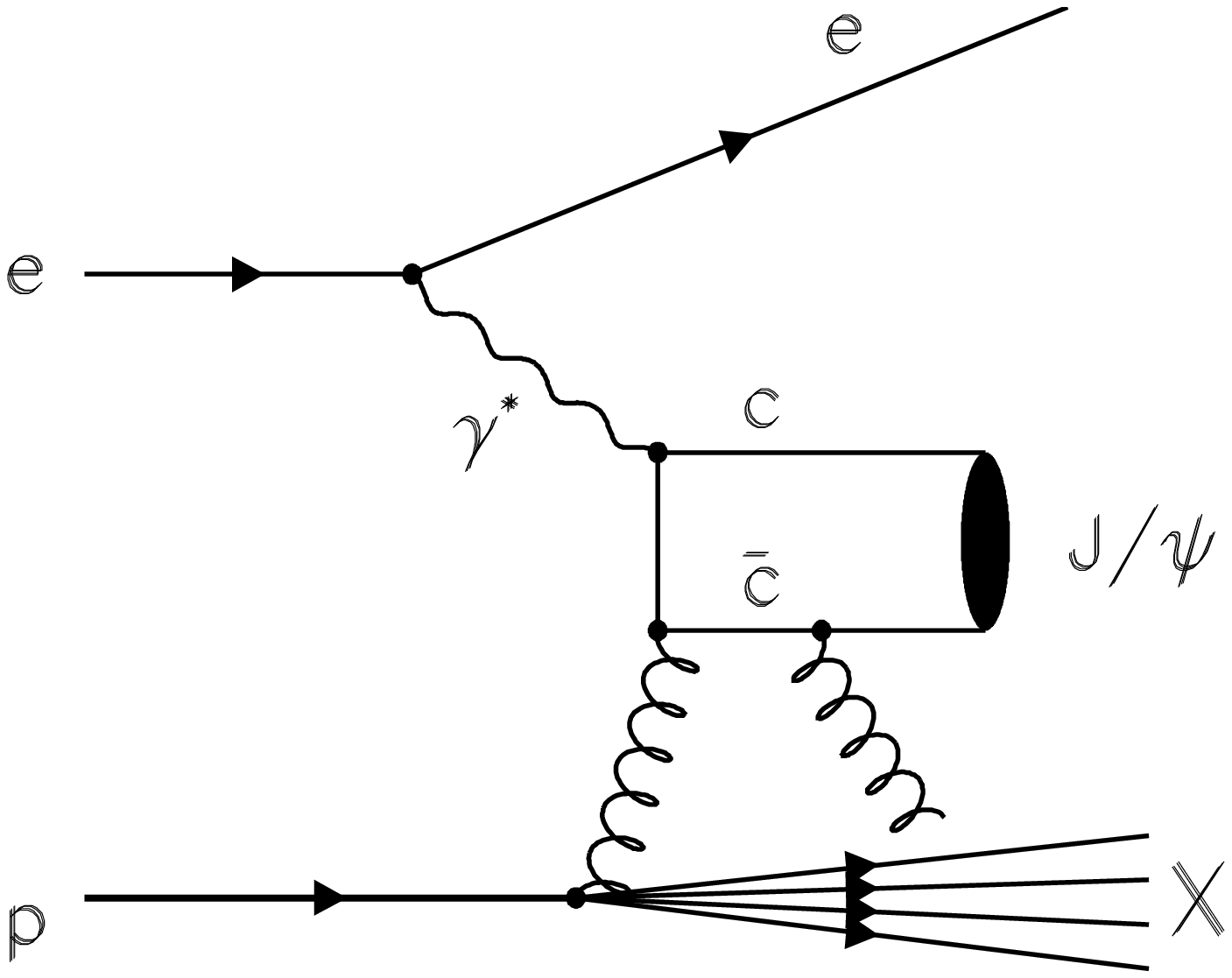,width=4.5cm}}
   \put( -5,15){\epsfig{file=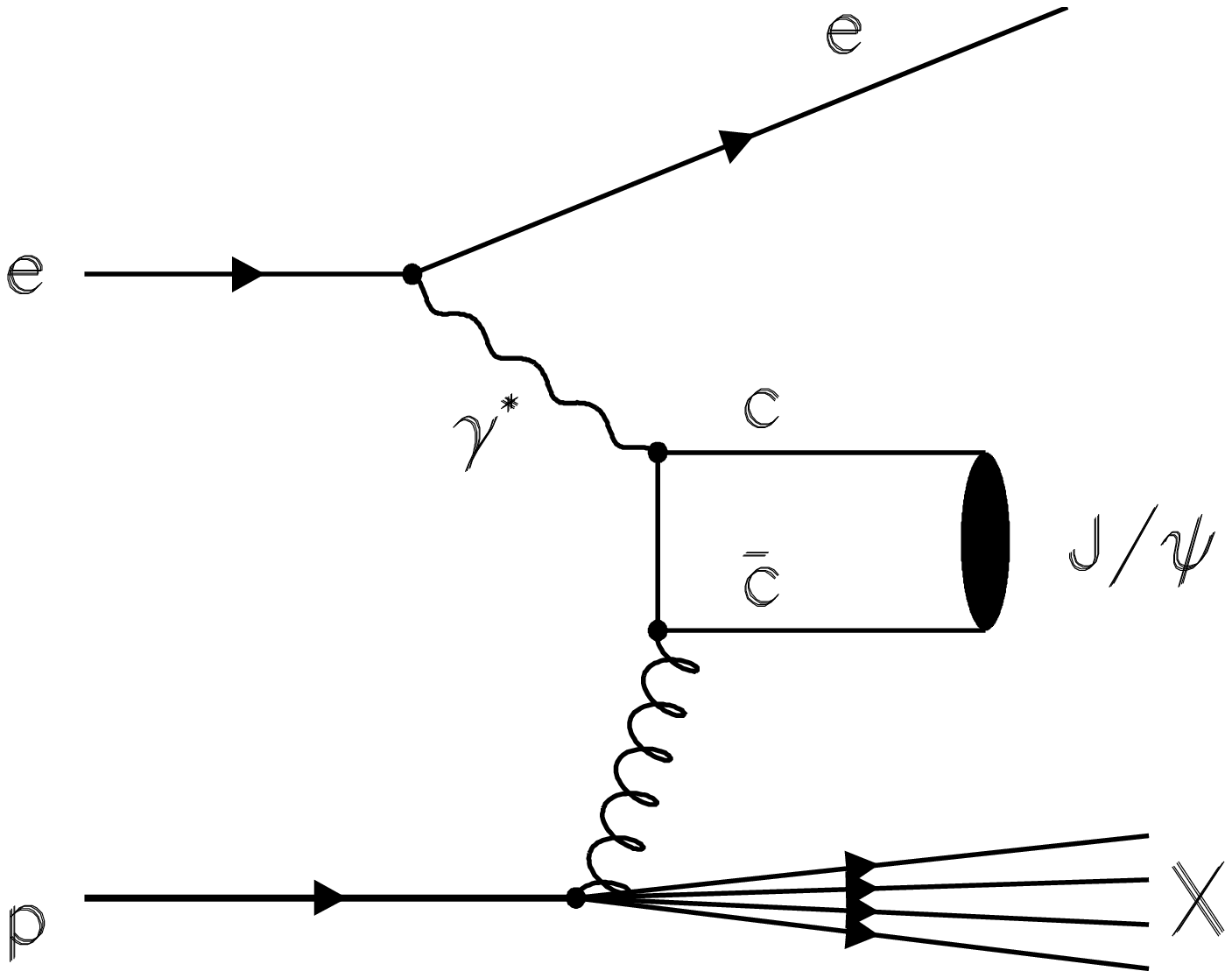,width=4.5cm}}
  \put(-3,80){or $\rho$}
  \put(35,83){or $\rho$}
 \end{picture}
 \end{center}
\vspace{-3.5cm}
\caption{\it Diagram for vector meson production.}
\label{fig:feynjpsi}
\vspace{-1.0cm}
\end{figure}
HERA allows to study the 
simplest photon-hadron diffractive reaction
$\gamma^* p \to {\rm VM} \; p$ in a new range of $\gamma^* p$
centre of mass energies up to $\Wgp \lsim 300$ \GeVx.
It is possible to investigate the transition from the soft
regime of photoproduction of light vector mesons at low
momentum transfers to the pQCD regime in the presence
of three hard scales: $Q^2$, $M_X=M_V$ and $t$.
The traditional way of modelling vector meson production
is the vector meson dominance model where the
photon fluctuates into a vector meson which
then elastically scatters off the proton.
In such a picture  only a weak energy dependence of the cross section
is expected~\cite{th:dola2} (see Fig.~\ref{fig:feynjpsi} 
upper left). This works well for light vector mesons
at low \Qsq and $t$.
For heavier vector meson like the \jpsi~the charm mass
provides a hard scale and pQCD calculations are possible.
An example for such a pQCD model where the \jpsi~couples
via two gluons to the proton
is shown in Fig.~\ref{fig:feynjpsi} (upper right).

\begin{figure}[h] \unitlength 1mm
 \vspace{-3.0cm}
  \begin{center}
  \begin{picture}(0,100)
    \put(-40,0){\epsfig{file=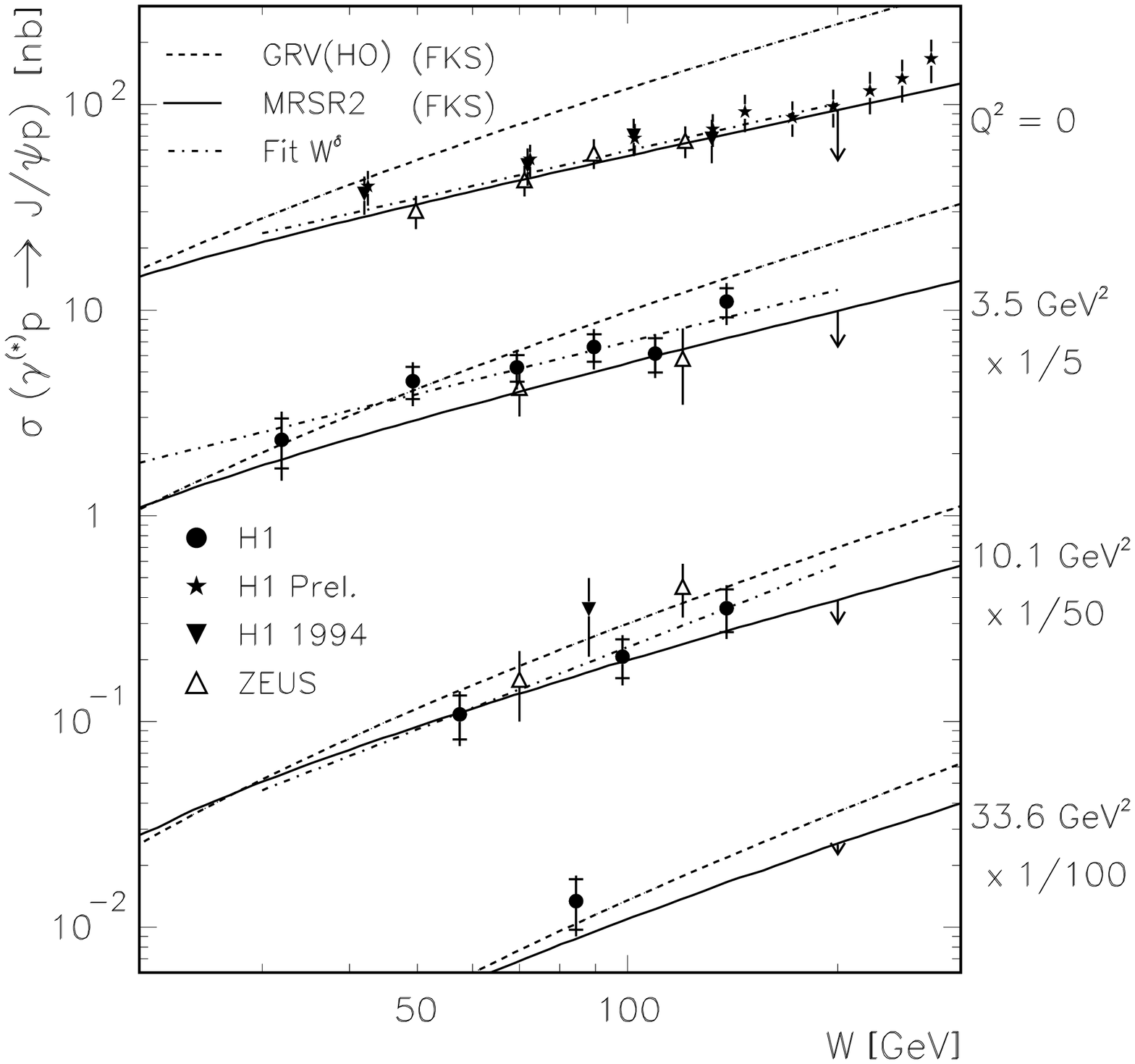,width=8.cm}}
    \put(25,65){$\delta = 0.77 \pm 0.18$ }
    \put(25,43){$\delta = 0.84 \pm 0.2$ }
    \put(25,24){$\delta = 1.3 \pm 0.4$ }
  \end{picture}
 \vspace{-1.3cm}
 \caption{\it The elastic \jpsi ~cross section 
 as a function of \Wgp.}
 \label{fig:jpsivsw}
  \end{center}
\vspace{-1.0cm}
 \end{figure}

The measured elastic \jpsi~cross section 
as a function of \Wgp ~is displayed
in Fig.~\ref{fig:jpsivsw} for 
$0 < \Qsq < 33$ \GeVsqx. 
A wide \Wgp~ range from $50 < \Wgp < 285$ \GeV is
covered. A much steeper rise ($\Wgp^{\! \! 0.77 \pm 0.18}$)
than predicted by soft models ($\Wgp^{\! \! 0.32}$)
is observed in the data. However, the data are
well described by a QCD inspired calculation~\cite{fks}.
The arrow indicates the uncertainty introduced 
by the charm mass ($m_c=1.4-1.5$ \GeVx).
Since the gluon density enters squared in the calculation
this measurement is sensitive to the gluon distribution
in the proton.
The cross section decreases as $1/{(\Qsq + M_{\sjpsi}^2)}^n$,
where $n=2.38 \pm 0.11$, as \Qsq increases.
The data are not precise enough to see
a further steepening of the rise with increasing \Qsqx.
In conclusion,
VDM holds unless there is a hard scale in the problem.
The failure of this picture together with the sucess of
QCD inspired models gives additional confidence in the
potential of QCD to describe processes where the scales
involved are surprisingly small. Further studies will
gain a better understanding of the transition from the
region governed by soft interactions to the regime where
pQCD turns on.
%


%
 %
%
\begin{figure}[h] \unitlength 1mm
\vspace{-2.3cm}
 \begin{center}
 \begin{picture}(0,100)
   \put(-35,0){\epsfig{file=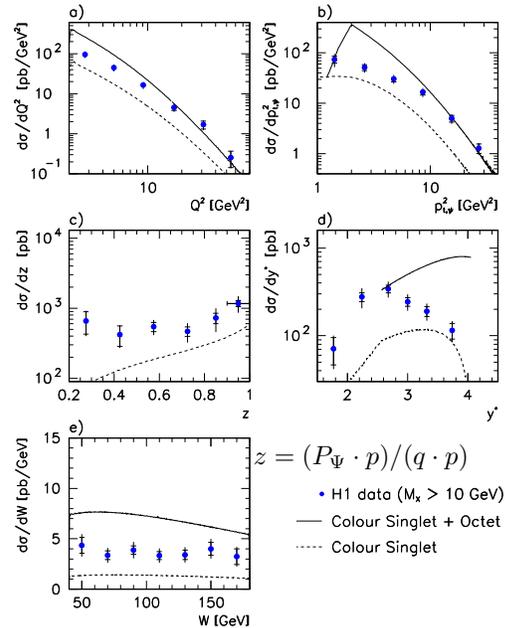,
   width=7.5cm}}
   \put(0,26){$z = (P_{\scriptsize \Psi} \cdot p)/( q \cdot p)$ }
 \end{picture}
 \end{center}
\vspace{-1.5cm}
\caption{\it Differential cross section for inelastic \jpsi~
production as a function of \Qsqx, $P^2_T$, $z$, \Wgp~ and the
rapidity.}
 \label{fig:jpsiinel}
\vspace{-1.0cm}
\end{figure}
%
Inelastic vector meson production is characterised
by an inelasticity
$ 0 < z = (P_{\scriptsize \Psi} \cdot p)/( q \cdot p) < 0.9$ 
and a relatively large mass $M_X$.
Two models for the production meachnism are shown
in the lower part of Fig.~\ref{fig:feynjpsi}.
In a short distance process a
$c \bar{c}$ state is produced in a boson-gluon fusion.
The  $c \bar{c}$ system is therefore a colour octet. 
The long range transition into the \jpsi~ has therefore to involve
the emission of an additional gluon.
The failure of this approach to describe the production of
quarkonia at large $P_T$, in particular at TEVATRON, 
led to the inclusion of additional
contributions where the $c \bar{c}$ system transforms
into an colour singlet in a long range interaction
without emitting a gluon~\cite{braaten}. The long range
matrix elements have been fitted to the $p\bar{p}$ data.
It is therefore interesting to see, if these fits
can be sucessfully applied to HERA data.

The differential cross section for inelastic \jpsi~ production
defined by $M_X > 10$ \GeV is shown in Fig.~\ref{fig:jpsiinel}.
The colour octet model is overlayed as dashed the singlet
model as solid line. Both predictions~\cite{fleming}
fail to describe the shape of the data. 
The colour singlet is a factor $2-3$ below the data,
the sum of the singlet and octet model is too large.
The reason for this failure is unknown.
It seems that an
overall adjustment of transition matrix element is necessary.
The failure to describe the shape also calls for an
relative adjustment of individudal contributions.
Recently it has been reported that the inclusion of 
higher order radiation (in a Monte Carlo approach)
lowers the colour octet matrix elements needed to describe
the TEVATRON data~\cite{Cano-Coloma}. Similar effects might also
be important at HERA.

%% file: conclusions.tex
In the past years HERA has been operated very successfully
and provided useful $e^\pm p$ data sets.
From the recent $1998/1999$ $e^- p$ running period
neutral and charged current DIS cross section have
been measured. The observed difference in the 
$e^- p$ and $e^+ p$ data provides clear evidence for 
electroweak interference effects.
HERA starts to probe the electroweak sector of the Standard Model.
So far, no striking deviations have been seen, although
for $\Qsq > 15000 \GeVsq$ more events than expected are found in the
$e^+ p$ data. 
Future high precision data will reveal, if this is only a statistical
fluctuation or if it is a first hint for a new interaction on top of QCD.
The HERA collider will be upgraded in the year $2000$ to deliver an integrated
luminosity of $150$~\pbinv ~per year. This will allow to accumulate
$\int {\cal L} dt \approx 1 \; {\rm fb}^{-1}$ until the year $2006$. 

Precise data on the proton structure function constrain the parton density
functions in the proton. This inclusive measurement is consistent with
the parton evolution according to the DGLAP equations.
The gluon density has been directly measured from observables based on 
the hadronic final state like dijet and charm production.
One of the biggest uncertainties in these measurement is the unknown
relation between the partonic final state and the measurable hadrons.
First steps to develop analytical calculations for specific
observables based on  the hadronic final state in the current hemisphere
of the Breit system are very encouraging. 
Further theoretical and experimental work is needed here.

New tests of forward particle production put the DGLAP evolution scheme
into question. More partonic activity seems to be required. Moreover it
has been observed that the probability to emit hard partons in the forward 
region is constant with respect to the inclusive cross section. 
This is very similar to the behaviour of event where no particles are emitted
in a rapidity region.

So far HERA data have supported a consistent picture of the production mechanism
of rapidity gap events. New results on diffractive charm production
seem to be in disagreement with the pomeron picture. More data are needed
to clarify this situation.

Precise data on vector meson production are now available.
In the elastic channel QCD models are able to describe the data.
However, in the inelastic channel no satisfying calculation is available.